\documentclass[11pt,letter]{article}
\pdfoutput=1 

\usepackage{jheppub}
\usepackage{amssymb,amsmath,amsthm,graphicx}
\usepackage{graphics, color}
\usepackage{subfigure}
\usepackage{latexsym}
\usepackage{bm}
\usepackage[hang,flushmargin]{footmisc}
\usepackage{epsfig}
\usepackage{textcomp}
\usepackage{cancel}
\hypersetup{
 pdfauthor={Gary T. Horowitz, Maciej Kolanowski, Jorge  E. Santos},
 citecolor=black,linkcolor=black,urlcolor=black}
\usepackage[utf8]{inputenc}
\allowdisplaybreaks[1]

\newcommand{\be}{\begin{equation}}
\newcommand{\ee}{\end{equation}}
\newcommand{\DD}{\Delta\!\!\!\!\Delta}
\newcommand{\Grad}{\nabla\!\!\!\!\nabla}

\begin{document}
\title{Frozen Firewall: Generic Singularity Formation on an Extremal Horizon}

\author[1]{Gary~T.~Horowitz,}
\author[1]{Maciej~Kolanowski}
\author[2]{and Jorge~E.~Santos}

\affiliation[1]{Department of Physics, University of California, Santa Barbara, CA 93106, U.S.A.}
\affiliation[2]{DAMTP, Centre for Mathematical Sciences, University of Cambridge, Wilberforce Road, \\ Cambridge CB3 0WA, UK}
\emailAdd{horowitz@ucsb.edu}
\emailAdd{mkolanowski@ucsb.edu}
\emailAdd{jss55@cam.ac.uk}

\newcommand{\blue}{\color{blue}}
\newcommand{\red}{\color{red}}

\abstract{

It is known that linearized perturbations of extremal black holes result in growing curvature on the horizon. However, nonlinear perturbations typically do not evolve to extremal black holes and do not have growing curvature at late times. We show that a large class of nonlinear perturbations of an extremal planar anti-de Sitter black hole does have horizon curvature that grows unbounded in time.   The late time behavior of the nonlinear evolution is found to be captured by a linearized analysis. We argue that the generic nonlinear perturbation behaves similarly. }
\maketitle
\section{Introduction}

Aretakis showed more than ten years ago that extremal black holes are unstable to linear perturbations \cite{Aretakis:2011ha,Aretakis:2011hc,Aretakis:2012ei}. The value of a massless scalar field on an extremal horizon decays in time, but its first transversal derivative does not, and higher derivatives blow up. Soon after, it was shown that linear Einstein-Maxwell perturbations result in certain components of the curvature approaching a constant \cite{Lucietti:2012xr,Apetroaie:2022rew} or even growing along the horizon \cite{Gralla:2016sxp}. These correspond to tidal forces felt by ingoing observers.  In contrast, nonlinear perturbations of extremal black holes typically do not result in growing curvature. The solutions either settle down to nonextremal black holes, or do not form any black hole at all.  One must fine tune one parameter in the initial data to approach an extremal solution \cite{Murata:2013daa,Angelopoulos:2024yev}.

The above statement about nonlinear perturbations is true for compact horizons (like the familiar spherical black holes) but not in general. In anti-de Sitter (AdS) spacetime, there are noncompact horizons such as planar and hyperbolic black holes. They have finite energy density and charge density, but infinite horizon area. Finite energy deformations cannot approach  nonextremal planar black holes since they differ by an infinite amount of energy. So generic deformations will evolve toward the extremal solution. This is a perfect setting to study how nonlinearities affect the linearized instability. Does the growing curvature get suppressed by the nonlinearities or enhanced? Could the nonlinear evolution result in singularities forming in finite time on the horizon?

In this paper we study linear and nonlinear perturbations of the extremal planar AdS black hole. At the linear level, we find that tidal forces on infalling observers grow at a rate depending on the wave vector $k$ along the planar horizon. This is true in all dimensions $D \ge 4$, but there is an interesting difference between $D=4$ and $D>4$. In  $D=4$, the fastest growth occurs at $k=0$, while in higher dimensions, the fastest growth occurs for $k= k_0 \ne 0$, where $k_0$ depends on the charge density, cosmological constant, and $D$.  We then study the  nonlinear evolution by numerically solving the Einstein-Maxwell equations. We will focus on $D=5$ both to see the effects of a nonzero $k_0$, and also for numerical convenience.  At late times, we find tidal forces growing on the horizon with a profile dominated by the $k=k_0$ linear mode, and at a rate given by the linear theory. Our numerical integration is restricted to initial data preserving an $SO(3)$ symmetry, and we see this same late time behavior for all initial deformations in this class. Only the overall amplitude depends on the initial conditions. In other words, the late time behavior of the curvature in the full nonlinear evolution is  captured by the linear theory.

It is perhaps disappointing that the full nonlinear evolution does not enhance the linear instability of the extremal horizon and cause singularities to form in finite time. However in our case there is an intuitive explanation why the late time behavior can be described by a linear analysis. We are adding a localized finite energy excitation to an infinite planar black hole. As time evolves, the excitation spreads out over an increasing area with decreasing amplitude.  Eventually, the amplitude is small enough that a linear analysis is applicable.\footnote{This may not be the complete explanation since the late time behavior of a nonlinear perturbation of an extremal charged black hole generated by a (finely tuned) spherical scalar field  also resembles the linear Aretakis instability \cite{Angelopoulos:2024yev}.} This intuitive picture strongly suggests that all nonlinear deformations (even those breaking the $SO(3)$ symmetry) will have the same late time behavior.  Even though the amplitude is decreasing, its gradient off the horizon grows, so that the curvature on the horizon actually increases.  The fact that a linearized analysis can correctly describe an exact solution with large curvature is possible in general relativity, and occurs, e.g., in exact gravitational plane waves with diverging amplitude.

The fact that we generically find growing curvature on the horizon is reminiscent of  a firewall, which is a singularity on the horizon that is conjectured to form when a black hole evaporates \cite{Almheiri:2012rt}. 
 Since our singularity only forms on zero temperature horizons, one might call it  a ``frozen firewall". 

Since the curvature is growing without bound on the horizon, one might expect that it also becomes large a small affine distance away. This would violate the spirit of (weak) cosmic censorship since generic evolution would produce arbitrarily large curvature that is visible from infinity. Unfortunately, this does not happen. The growing curvature is concentrated closer and closer to the horizon, and the curvature decays at any fixed affine distance away.

The next section contains our linear analysis. After first reviewing our background spacetime, we discuss a test scalar field as a simple toy model of what to expect, and then examine the $D=5$ linear Einstein-Maxwell equations with $SO(3)$ symmetry. (Appendix \ref{app:scalar} contains some results of the numerical evolution of the scalar field, and a more complete discussion of Einstein-Maxwell perturbations in arbitrary dimensions without any symmetry restriction is given in Appendix \ref{app:B}.)
In Section \ref{sec:nonl}, we describe the initial data and the procedure for numerically evolving the full nonlinear equations (with some further details in Appendices \ref{app:pol}, \ref{app:lagrange}, and \ref{app:appbessel}.) Section \ref{sec:results} contains our results for both the linear and nonlinear evolutions, and  the final section contains some concluding remarks.

\section{Linear  analysis }

In this section we consider linear perturbations.
 To fix our conventions, we will start with a short review of the (extremal) planar Reissner-Nordstr\"om (RN) solution. Then, we shall move to a toy model of a Klein-Gordon field which already reveals key features  underlying the Aretakis instability. Having done this, we will perform a mode analysis for the Einstein-Maxwell system, showing that these features appear there as well (albeit in a stronger version). 

\subsection{Background}
We start with the Einstein-Maxwell action in $D= n+2$ dimensions:
\begin{equation}
    S = \int d^{n+2}  x \sqrt{-g} \left(
R - 2 \Lambda - \frac{1}{4}F_{ab} F^{ab}
    \right)\,,
\end{equation}
where $R$ is the $(n+2)$-dimensional Ricci scalar of the bulk spacetime metric $g$, 
$\Lambda$ is a negative cosmological constant, and $F = \mathrm{d}A$.

The equations of motion are as follows:
\begin{subequations}
    \begin{equation}
        R_{ab} - \frac{R}{2}g_{ab} + \Lambda g_{ab} = \frac{1}{2} \left(
F_{a}^{\ \ c} F_{bc} - \frac{g_{ab}}{4} F_{cd}F^{cd}
        \right)
    \end{equation}
    and
    \begin{equation}
        \nabla^a F_{ab} = 0.
    \end{equation}
\end{subequations}%
Homogeneous black holes can have spherical, planar, or hyperbolic horizons depending on the asymptotically AdS boundary conditions.  We will be interested in planar black holes, which are given by
\begin{subequations}
    \begin{equation}
        {\rm d}s^2 = -f {\rm d}t^2 + \frac{{\rm d}r^2}{f} + r^2 \delta_{ij} {\rm d}x^i {\rm d}x^j,
    \end{equation}
    \begin{equation}
        A = \sqrt{\frac{2n}{n-1}} Q r^{1-n} {\rm d}t\,,
    \end{equation}
    where 
    \begin{equation}
        f = \frac{r^2} {L^2} - \frac{2M}{r^{n-1}} + \frac{Q^2}{r^{2n-2}},
    \end{equation}
    and 
    \begin{equation}
       \frac{1}{L^2} = 
      - \frac{2\Lambda}{n(n+1)} .
    \end{equation}
\end{subequations}%
 The event horizon is located at the largest root of $f$, denoted $r_+$. When $r_+$ is a double root, the black hole is extremal. In this case, the parameters $M$ and $Q$, which determine the mass density and charge density, are related to $r_+$ by
\begin{subequations}
    \begin{equation}
        M= \frac{  n r_+^{n+1}}{(n-1)L^2},
    \end{equation}
    \begin{equation}
        Q = \pm \frac{r_+^n}{L} \sqrt{ \frac{n+1}{n-1}}.
    \end{equation}
\end{subequations}%
At extremality, we may take the near-horizon limit to obtain
\begin{subequations}
\begin{equation}
    {\rm d}s^2 = -\frac{n(n+1)  \rho^2}{L^2} {\rm d}\tilde{t}^2 + \frac{{L^2 \rm d}\rho^2}{n(n+1)  \rho^2} + r_+^2 \delta_{ij}{\rm d}x^i {\rm d}x^j,
\end{equation}
\begin{equation}
    F = \sqrt{2 n (n-1)} \frac{Q}{r_+^{n}} {\rm d}\tilde{t} \wedge {\rm d}\rho.
\end{equation}
\end{subequations}%

Since we are mainly interested in the behavior of various fields {\it at} the horizon, it is convenient to work with ingoing Eddington--Finkelstein coordinates in which the original metric is
\begin{equation}\label{eq:EFform}
    {\rm d}s^2 = -f {\rm d}v^2 + 2 {\rm d}v {\rm d}r + r^2 \delta_{ij} {\rm d}x^i {\rm d}x^j.
\end{equation}
The near-horizon metric analogously is 
\begin{equation}
    {\rm d}s^2 = -\frac{n(n+1)  \rho^2}{L^2} {\rm d}\tilde{v}^2 + 2{\rm d}\tilde{v} {\rm d}\rho + r_+^2 \delta_{ij} {\rm d}x^i {\rm d}x^j.
\end{equation}
Note that the near-horizon geometry has enhanced symmetry with respect to the full geometry. The symmetry group is $O(2,1) \times \mathbb{R}^3$, instead of $\mathbb{R}_v \times \mathbb{R}^3$. Of particular importance for our discussion is the connected component of the identity, $\mathrm{SO}_0(2,1) \subset \mathrm{O}(2,1)$, generated by exponentiating the Lie algebra elements $L_{-1}$, $L_{0}$, and $L_{1}$, given by
\begin{equation}
\begin{aligned}
&L_{-1}=\partial_{\tilde{v}}\,,
\\
&L_{0}=\tilde{v}\partial_{\tilde{v}}-\rho \partial_{\rho}\,,
\\
&L_{1}=\tilde{v}^2\partial_{\tilde{v}}-2\left[\frac{L^2}{n(n+1)}+\tilde{v}\rho\right]\partial_{\rho}\,,
\end{aligned}
\end{equation}
which parametrise time translations, dilations, and special conformal transformations, respectively. These obey the usual Lie algebra brackets for $\mathfrak{so}(2,1)$:
\begin{equation}
[L_0,L_{-1}]=-L_{-1}\,,\quad [L_0,L_{1}]=L_{1}\quad \text{and}\quad [L_1,L_{-1}]=-2 L_0\,.
\end{equation}

\subsection{Toy model -- scalar field} \label{sec:toy_model}

In general, we are interested in solving the two-dimensional wave-like equation:
\begin{equation}
    \nabla^a \nabla_a u - W(r) u = 0, \label{eq:wavelike}
\end{equation}
where $u=u(v,r)$ but $\nabla_a$ is the spacetime $(n+2)$-dimensional covariant derivative. If we think about this equation as describing some $(n+2)$-dimensional fields, $u$ will be a component in the Fourier decomposition with respect to the $x^i$ directions and then, generically $W$ will depend on the momentum $k$. Let us denote $\mu^2 = W(r_+)$. For discrete choices of $\mu$ (but otherwise arbitrary $W$), one can show that a certain linear combination of $u$ and (a finite number of) its radial derivatives is conserved on the horizon,  $r = r_+$, in the spirit of \cite{Aretakis:2011ha,Aretakis:2011hc,Aretakis:2012ei}. As just discussed, generically $\mu$ will be a function of planar momenta $k$, thus these conservation laws can only occur for special components in the Fourier decomposition. Since the generic initial data will contain all possible momenta $k$, we will not pursue this further here and will treat $\mu$ as arbitrary.

For our background \eqref{eq:EFform}, the wave-like equation \eqref{eq:wavelike} explicitly reads
\begin{equation}
    f \partial_{rr}^2 u + 2 \partial^2_{vr} u + \frac{n}{r} \partial_v u + \frac{n f + r f'}{r} \partial_r u - W(r) u = 0.
\end{equation}
Taking the near-horizon, late-time limit, we get
\begin{equation}
    \frac{n(n+1)}{L^2} \rho^2 \partial^2_{\rho \rho} u + 2 \partial^2_{\tilde{v} \rho} u +\frac{2n (n+1)}{L^2} \rho \partial_\rho u - \mu^2 u = 0.
\end{equation}
Based on the symmetries of this equation, it was argued that the late-time dynamics should be governed by the following ansatz \cite{Gralla:2017lto, Gralla:2018xzo}:
\begin{equation}
    u(\tilde{v}, \rho) = \tilde{v}^{-\alpha} h(\rho \tilde{v}), \label{eq:ansatz_late_time}
\end{equation}
where $h$ is a regular function at the origin and decays sufficiently fast away from it. The parameter $\alpha$ must be determined from the equations of motion.
Before we do that, let us motivate this ansatz and explain why we expect that the equations of motion will admit solutions with this particular functional form. This is simply because the near-horizon background is preserved by dilations and this form of $u$  assumes that it is an eigenfunction of the dilation generator $L_0=\tilde{v} \partial_{\tilde{v}} - \rho \partial_\rho$. Without further ado, we find the following equation for $h$:
\begin{equation}
   X h''(X) \left(2+X\frac{  n^2+  n}{L^2}\right)+2 h'(X) \left(\alpha +X\frac{ n^2 + n }{L^2}+1\right)-\mu ^2 h(X) = 0.
\end{equation}
Solving this for $h$, we obtain the following solution for $u$:
\begin{subequations}
    \begin{equation}
        u = u_0 \tilde{v}^{-\alpha}
        \left(
        2 + \tilde{v} \rho \frac{n^2+n}{L^2}
        \right)^{-\alpha},
    \end{equation}
    where
    \begin{equation}\label{eq:alpha}
        \alpha = \frac{1}{2} + \frac{1}{2} \sqrt{1 + \frac{4\mu^2 L^2}{n(n+1)}}
    \end{equation}
    \label{eq:special}
\end{subequations}
and $u_0$ is a normalization constant.
A few comments are in order:
\begin{itemize}

   \item  Notice that $\alpha$ is just the conformal dimension of a scalar field with mass $\mu$ in the $\textrm{AdS}_2$ background of the $\textrm{AdS}_2 \times \mathbb{R}^n$ near-horizon geometry

    \item The function $u$ appearing in Eq.~(\ref{eq:special}) is annihilated by the special conformal generator, $\pounds_{L_1} u = 0$, and transforms as an eigenfunction of the dilation generator with eigenvalue $-\alpha$, i.e., $\pounds_{L_0} u = -\alpha\, u$. This means that $u$ is a conformal primary of weight $\alpha$ in AdS$_2$. The lowering operator $L_{-1}$ generates descendants from the primary function $u$. Explicitly, repeated action of $L_{-1}$ produces the tower of descendants $u, \quad \pounds_{L_{-1}} u, \quad \pounds_{L_{-1}}^2 u, \quad \dots,$ each of which is an eigenfunction of the dilation generator $L_0$ with eigenvalues decreasing by integers, 
\begin{equation}
\pounds_{L_0} \big(\pounds_{L_{-1}}^j u\big) = -(\alpha+j)\, \pounds_{L_{-1}}^j u, \quad j=0,1,2,\dots.
\end{equation}
This construction realizes a highest-weight representation of the $\mathfrak{sl}(2,\mathbb{R})$ algebra, with $u$ as the conformal primary and the $\pounds_{L_{-1}}^j u$ as its descendants. Note that because $L_{-1}$ is a Killing vector field, it commutes with $\Box_{{\rm AdS}_2}$, and as such each of the descendants is still a solution of the corresponding Klein-Gordon equation in AdS$_2$.
    \item The field $u$ will decay outside the horizon together with all its derivatives.
    \item The field will decay along the horizon but $\partial_\rho^{\lceil{\alpha} \rceil} u$ will blow up as $\tilde{v} \to \infty$, where $\lceil{\alpha} \rceil$ denotes the smallest integer containing $\alpha$.
    \item When $\alpha$ is a natural number, $\partial_\rho^{\alpha} u$ will approach a constant at late times. For these values of $\mu$, one can derive the aforementioned conservation laws.
    \item A priori, $\mu^2$ can have an arbitrary sign. Our analysis works  as long as $\alpha$ remains real. When it stops being real, $u$ fails to satisfy the two-dimensional BF bound which is a sign of instability in the full asymptotically $\textrm{AdS}$ geometry \cite{Hollands:2014lra}. In this case, a solution would develop  hair outside the horizon. 
    \item Eq. \eqref{eq:alpha} implies that $\alpha > 0$. Thus, there is no choice of parameter $\mu$ for which the solution (without taking any derivatives) would actually grow along the horizon.
    \item We may estimate that at late times, $\partial_{\rho}^{\lceil{\alpha} \rceil} u$ will be large not only on the horizon but also on a thin region of up to radius $\rho \sim \tilde{v}^{-1}$.
\end{itemize}

Let us finish by providing the simplest example: a free scalar field $\phi$ of mass $m^2$. We may decompose it into Fourier modes:
\begin{equation}
    \phi = \int \textrm{d}^n \vec{k}\,\phi_{\vec{k}} (v, r) e^{i\vec{k}\cdot \vec{x}}.
\end{equation}
Then, each component satisfy
\begin{equation}
    \nabla^a \nabla_a \phi_{\vec{k}} - \left(
        m^2 + \frac{k^2}{r^2}
    \right)\phi_{\vec{k}} = 0\,,
\end{equation}
with $|\vec{k}|\equiv k$. In particular, we find\footnote{One should notice that in our conventions $x$ is dimensionless and so is $k$. Thus, $\frac{k^2}{r_+^2}$ has units of mass squared.} $\mu^2 = m^2 + \frac{k^2}{r_+^2}$. It follows that the smallest $\alpha$ (and thus, the largest growth after taking $N> \alpha$ derivatives) corresponds to the homogeneous $k=0$ perturbation.

To be more precise, if we assume that the ansatz \eqref{eq:ansatz_late_time} holds, we may write the $p$-derivative of our field on the horizon
\begin{equation}
    \partial^p_{\rho}\phi(\tilde{v},\rho=0,\vec{x}) = \int \mathrm{d}^n \vec{k}\,\tilde{\phi}_{\vec{k}}\,\tilde{v}^{p}\,\tilde{v}^{-\alpha(k)}e^{i \vec{k}\cdot\vec{x}} = \tilde{v}^{p}\int \mathrm{d}^n \vec{k}\,\tilde{\phi}_{\vec{k}}\,e^{-\alpha(k)\,\log \tilde{v} + i \vec{k}\cdot\vec{x}},
\end{equation}
where $\tilde{\phi}_{\vec{k}}$ is an amplitude that we take to be a smooth function of $\vec{k}$. Since we are interested in the late-time behavior $(\tilde{v} \to \infty)$, we can evaluate it using a saddle-point approximation. We may write the field as
\begin{equation}
    \partial_\rho^p\phi(\tilde{v},\rho=0,\vec{x}) = \tilde{v}^{p}\int \mathrm{d}\Omega_{n-1} \int k^{n-1} \mathrm{d}k\,\tilde{\phi}_{\vec{k}}\,e^{-\alpha(k)\,\log \tilde{v}+i \vec{k} \cdot \vec{x}},
    \label{eq:near}
\end{equation}
where $\mathrm{d}\Omega_{n-1}$ is the volume element (or volume form) of the $(n-1)$-dimensional unit sphere. We see that only the last integral can be evaluated in the saddle-point approximation since 
 the large parameter $\log \tilde v$ only multiplies a function of $k$. We have
\begin{equation}
    \alpha(k) = \frac{1}{2} + \frac{1}{2} \sqrt{1 + \frac{4 L^2}{n(n+1)}\left(m^2 + \frac{k^2}{r_+^2} \right) }.
    \label{eq:alphak}
\end{equation}
This function has only one minimum at $k = 0$, around which we can write
\begin{equation}
    \alpha(k) = \frac{1}{2} \left[\sqrt{1+\frac{4 L^2 m^2}{n(n+1)}}+1\right]+\frac{k^2 L^2}{r_+^2 \sqrt{n (n+1) \left(4 L^2 m^2+n^2+n\right)}}+\mathcal{O}\left(k^4\right)
\end{equation}
and so
\begin{equation}
    \partial_\rho^p\phi(\tilde{v},\rho=0,\vec{x}) \approx \frac{2\pi^{n/2}}{\Gamma(n/2)} \tilde{\phi}_0 \tilde{v}^{p-\alpha(0)} \int_0^\infty k^{n-1} \mathrm{d}k\,e^{-\frac{k^2}{2}\alpha''(0)\log \tilde{v}} =\left[\frac{2 \pi }{\alpha ''(0)}\right]^{n/2}\frac{ \tilde{v}^{p-\alpha (0)}}{\log ^{\frac{n}{2}}\tilde{v}}  \tilde{\phi} _0\,.
\label{eq:focus}
\end{equation}
The argument above breaks down when $\alpha(0) = \frac{1}{2}$, corresponding to $4L^2 m^2 = -n(n+1)$, in which case 
\begin{equation}
    \alpha(k) = \frac{1}{2} + \frac{L}{r_+ \sqrt{n(n+1)}} k.
\end{equation}
This mass is still allowed, since it lies above the Breitenlohner-Freedman bound in AdS$_{n+2}$. As a linear function, $\alpha$ does not have a saddle point. Instead it is dominated by the endpoint contribution at $k=0$ which means that
\begin{equation}
\partial_\rho^p\phi(\tilde{v},\rho=0,\vec{x}) \propto \frac{\tilde{v}^{p-\frac{1}{2}}}{\log ^{n}\tilde{v}}\,.
\label{eq:saddlepoinex}
\end{equation}
This behavior is characteristic of scalar fields that are on the verge of becoming unstable. If $4L^2m^2 < -n(n+1)$, the corresponding effective mass from the AdS$_2$ perspective falls below the AdS$_2$ Breitenlohner–Freedman bound, leading to an instability \cite{Durkee:2010ea,Hollands:2014lra} of the full extremal geometry.

 The solution given in Eq.~(\ref{eq:special}), with $\alpha$ defined as in Eq.~(\ref{eq:alphak}), is intended to describe a generic fixed-$k$ solution near the horizon for times $\tilde{v} \geq \tilde{v}_0 \gg 1$. The special solution in Eq.~(\ref{eq:special}) is nonzero on the horizon for all $\tilde{v}$, and one might wonder whether our late-time estimates would change if we instead considered initial data satisfying $\Phi(\tilde{v}_0,0,\vec{x}) = 0$ at $\tilde{v} = \tilde{v}_0$. Clearly, such initial data cannot be constructed via Fourier transforming the solution given in Eq.~(\ref{eq:special}). Indeed, one can go further and ask what happens if $j$ derivatives of the initial data vanish at $\tilde{v} = \tilde{v}_0$. In order to address this problem, we recall that we can generate new solutions by acting with $\pounds_{L_{-1}}$ on Eq.~(\ref{eq:special}). We can then construct a new solution $\tilde{u}$ as follows:
\begin{equation}
\begin{aligned}
\tilde{u}(\tilde{v},\rho)&=\left(\pounds_{L_-1}-\frac{1}{\tilde{v}_0}\right)u
\\
&=\frac{1}{\tilde{v} \tilde{v}_0 \left[2+\frac{n (n+1) \rho \tilde{v}}{L^2}\right]}\left[\frac{n (n+1) \left(\tilde{v}-\tilde{v}_0\right)^2\rho  }{L^2}+2 \left(\tilde{v}-\tilde{v}_0\right)-\frac{n (n+1)\,\rho \,\tilde{v}_0^2}{L^2}\right]u(\tilde{v},\rho)\,,
\end{aligned}
\end{equation}
which vanishes linearly in $\rho$ at $\tilde{v} = \tilde{v}_0$, as desired. The late-time behavior of this solution can be readily understood by applying $\left(\pounds_{L_{-1}} - \tilde{v}_0^{-1}\right)$ to Eq.~(\ref{eq:focus}), which leads to the same late-time behavior due to the term proportional to $\tilde{v}_0^{-1}$.

One wishes to generalize this construction to data whose first $j$ derivatives with respect to $\rho$ vanish at $\tilde{v} = \tilde{v}_0$.
 The idea is just to take an appropriate linear combination of the solutions obtained by applying $\pounds_{L_{-1}}$ up to $j$ times. The operator we need to apply is now more complicated, but can be found in closed form, namely
\begin{equation}
\begin{aligned}
\tilde{u}(\tilde{v},\rho)&=\sum _{i=0}^j \left[1+\delta_i^j \left(\frac{1}{\alpha }-1\right)\right] \frac{\prod _{p=1}^{j-1-i} (\alpha -p)}{\prod _{p=1}^{j-1} (\alpha +p)} \frac{ \Gamma (1+j)}{ \Gamma(i+1) \Gamma (j+1-i)}\left.\frac{\partial^iu(\hat{v},\rho)}{\partial\hat{v}^i}\right|_{\hat{v}=\tilde{v}-\tilde{v}_0+1}
   \\
   &=\sum_{i=0}^j \frac{\Gamma (\alpha ) \Gamma (\alpha +1) \Gamma (j+1)}{\Gamma (i+1) \Gamma (j+1-i) \Gamma (j+\alpha ) \Gamma (i-j+\alpha +1)}\left.\frac{\partial^iu(\hat{v},\rho)}{\partial\hat{v}^i}\right|_{\hat{v}=\tilde{v}-\tilde{v}_0+1}\,.
\end{aligned}
\label{eq:firstj}
\end{equation}
The final equality holds for any $\alpha$, but special care is needed when interpreting it if $\alpha$ is an integer less or equal than $j$, in which case the first equality shows that all terms with $j -\alpha - 1-i \in \mathbb{N}^0$ vanish. We can now multiply the above expression by a $\vec{k}$-dependent function, integrate over $\vec{k}$, and then apply the saddle-point approximation. The term with the $i=0$ will always survive, and yield the same contribution as in Eq.~(\ref{eq:focus}). The special values of $\alpha$, and hence of $k$, for which some of the contributions above are absent, were previously identified in \cite{Lucietti:2012xr} as being associated with certain conserved charges. The argument above shows that, once inhomogeneities are included, these conserved charges are of limited use in controlling the generic late-time behavior of scalar linear perturbations along the horizon. In Appendix~\ref{app:scalar}, we perform numerical simulations of a test scalar field with $n = 3$ and $m^2 L^2 = -3$ in the full asymptotically AdS$_5$ spacetime,  and show that their late time behavior is indeed reproduced by the above near horizon analysis.

As we will see, the analysis is slightly different for the (linearized) Einstein-Maxwell system in dimension higher than four. To be more precise, the evolution can be reduced to master equations of the form \eqref{eq:wavelike} but with a more complicated potential, resulting in $\alpha$ having a (global) minimum at $|k_0| \neq 0$. Still, the  integral representation is the same:
\begin{equation}
\partial^p_{\rho}\phi(\tilde{v},\rho=0,\vec{x}) = \int \mathrm{d}^n \vec{k}\,\tilde{\phi}_{\vec{k}}\,\tilde{v}^{p}\,\tilde{v}^{-\alpha(k)}e^{i \vec{k}\cdot\vec{x}} = \tilde{v}^{p}\int \mathrm{d}^n \vec{k}\,\tilde{\phi}_{\vec{k}}\,e^{-\alpha(k)\,\log \tilde{v} + i \vec{k}\cdot\vec{x}},
\end{equation}
but the saddle-point approximation instead will yield
\begin{align}
\begin{split}
    \partial^p_\rho (\tilde{v},\rho=0,\vec{x}) &\approx \tilde{v}^{p-\alpha(k_0)} \int \textrm{d} \Omega_{n-1} \tilde{\phi}_{\vec{k_0}} k_0^{n-1} e^{i \vec{k_0} \cdot \vec{x}} \int_{\mathbb{R}}\textrm{d}k e^{-\frac{(k-k_0)^2}{2}\alpha''(k_0) \log \tilde{v}} \\ &= \sqrt{\pi} \frac{\tilde{v}^{p-\alpha(k_0)}}{\log^{1/2} \tilde{v}} k_0^{n-1} \int \textrm{d} \Omega_{n-1} \tilde{\phi}_{\vec{k_0}} e^{i \vec{k_0} \cdot \vec{x}}.
\end{split}
\end{align}
Thus, the suppression by $\log\tilde{v}$ is smaller and there is also a certain spatial modulation that emerges at late times.
\subsection{Einstein-Maxwell modes in $D=5$ with $SO(3)$ symmetry\label{subsec:em}}
We now consider linearized Einstein-Maxwell perturbations in  $D=5$. We will use Eddington–Finkelstein coordinates and deform the background $\mathbb{R}^3$ while preserving a round $S^2$, i.e. the linear analysis of this section will preserve $SO(3)$. (A general linear analysis in arbitrary dimensions is given in Appendix \ref{app:B}.) We write the background gauge field and metric as
\begin{equation}
\begin{aligned}
&{\rm d}s^2=-f(r){\rm d}v^2+2{\rm d}v\,{\rm d}r+r^2\left({\rm d}R^2+R^2{\rm d}\Omega_2^2\right)\,,
\\
&A=\frac{\sqrt{3}\,Q}{r^2}{\rm d}v\,,
\end{aligned}
\label{eq:back}
\end{equation}
where ${\rm d}\Omega_2^2$ denotes the line element of the unit round two-sphere, $R \in \mathbb{R}^+$ is the standard radial coordinate in spherical polar coordinates and
\begin{equation}
f(r)=\frac{r^2}{L^2}-\frac{2M}{r^2}+\frac{Q^2}{r^4}\,.
\end{equation}

Because we impose $SO(3)$ invariance, neither tensor-derived nor vector-derived perturbations are allowed. The only perturbations that contribute are scalar-derived gravito-electromagnetic modes. The natural harmonic on $\mathbb{R}^3$ under our restricted symmetry preserved analysis obeys
\begin{equation}
\frac{1}{R^2}\partial_R(R^2 \partial_R \mathbb{Y}_k)+k^2 \mathbb{Y}_k=0\,.
\end{equation}
Imposing regularity near the origin, yields
\begin{equation}
\mathbb{Y}_k(R)=\frac{\sin(k R)}{k R}\,,
\end{equation}
where we chose our normalisation so that $\mathbb{Y}_k(0)=1$. Our perturbations will thus be labeled by $k$, which is interpreted as the standard wavenumber of the corresponding gravito-electromagnetic wave.

Let $\Grad$ be the standard connection on $\mathbb{R}^3$, and $\DD$ the standard Laplacian. We then introduce the following two auxiliary quantities
\begin{equation}
\begin{aligned}
&\mathbb{S}^k_I=\Grad_I \mathbb{Y}_k
\\
&\mathbb{S}^k_{IJ}=\Grad_I\Grad_J \mathbb{Y}_k+\frac{\mathbb{G}_{IJ}}{3}k^2\mathbb{Y}_k\,,
\end{aligned}
\end{equation}
where $I,J$ range over the three spatial indices of $\mathbb{R}^3$, and $x^I$ denote the associated coordinate functions. We have also introduced $\mathbb{G}$ as the standard Euclidean metric on $\mathbb{R}^3$. Note that, by construction, $\mathbb{G}^{IJ}\mathbb{S}^k_{IJ}=0$.

Our metric and gauge field perturbations read 
\begin{equation}
\begin{aligned}
&\delta g=\left[h_{vv}(v,r){\rm d}v^2+h_{vr}(v,r){\rm d}v{\rm d}r+h_{rr}(v,r){\rm d}r^2\right]\mathbb{Y}_k(R)+2h_{v}(v,R)\mathbb{S}^k_I{\rm d}x^I{\rm d}v
\\
&\quad\quad\quad\quad\quad+2h_{r}(v,R)\mathbb{S}^k_I{\rm d}x^I{\rm d}r+h_{+}(v,r)\mathbb{Y}_k(R) \mathbb{G}_{IJ}{\rm d}x^I{\rm d}x^J+h_{-}(v,r)\mathbb{S}^k_{IJ}{\rm d}x^I{\rm d}x^J\,,
\\
&\delta A=\left[a_v(v,r){\rm d}v+a_r(v,r){\rm d}r\right]\mathbb{Y}_k(R)+a_R(v,r)\mathbb{S}^k_I{\rm d}x^I\,.
\end{aligned}
\end{equation}

There are ten functions of $v$ and $r$ in total that parametrise our perturbations. Up to this point, however, we have not made use of any gauge freedom. To incorporate it, we expand the infinitesimal diffeomorphisms $\xi$ in scalar harmonics $\mathbb{Y}_k$ as
\begin{equation}
\xi=\left[\xi_v(v,r){\rm d}v+\xi_r(v,r){\rm d}r\right]\mathbb{Y}_k(R)+\xi_R(v,r)\mathbb{S}^k_I{\rm d}x^I\,
\end{equation}
Similarly, we decompose the scalar that parametrises the standard $U(1)$ gauge freedom, $A \to A - {\rm d}\hat{\lambda}$, in the form
\begin{equation}
\hat{\lambda}=\lambda(v,r)\mathbb{Y}_k(R)\,.
\end{equation}

Under the transformations
\begin{equation}
\delta h_{ab}=\nabla_a\xi_b+\nabla_b\xi_a\quad\text{and}\quad \delta A_a=\xi^b\nabla_b A_a+A_b\nabla_a \xi^b-\nabla_a\lambda
\end{equation}
the  metric and gauge field perturbations transform as
\begin{subequations}
\begin{equation}
\begin{aligned}
&\delta h_{vv}(v,r)=-f(r) f'(r) \xi_r(v,r)-f'(r) \xi_v(v,r)+2 \partial_v\xi_v(v,r)\,,
\\
&\delta h_{vr}(v,r)=2 \left[f'(r) \xi_r(v,r)+\partial_r\xi_v(v,r)+\partial_v\xi_r(v,r)\right]\,,
\\
&\delta h_{rr}(v,r)=2 \partial_r\xi _r(v,r)\,,
\\
& \delta h_v(v,r)=\partial_v\xi_R{}(v,r)+\xi_v(v,r)\,,
\\
& \delta h_r(v,r)=-\frac{2 \xi _R(v,r)}{r}+\partial_r\xi _R(v,r)+\xi _r(v,r)\,,
\\
& \delta h_+(v,r)=2 r f(r) \xi _r(v,r)-\frac{2}{3} k^2 \xi _R(v,r)+2 r \xi _v(v,r)\,,
\\
& \delta h_-(v,r)=2 \xi _R(v,r)\,,
\end{aligned}
\end{equation}
and
\begin{equation}
\begin{aligned}
&\delta a_v(v,r)=-\frac{2 \sqrt{3} Q f(r)}{r^3} \xi_r(v,r)-\frac{2 \sqrt{3} Q}{r^3} \xi_v(v,r)+\frac{\sqrt{3} Q }{r^2}\partial_v\xi_r(v,r)-\partial_v\lambda(v,r)\,,
\\
&\delta a_r(v,r)=\frac{\sqrt{3} Q}{r^2} \partial_r\xi_r(v,r)-\partial_r\lambda(v,r)\,,
\\
&\delta a_R(v,r)=\frac{\sqrt{3} Q}{r^2} \xi _r(v,r)-\lambda(v,r)\,.
\end{aligned}
\end{equation}
\end{subequations}%
Since $\xi_v$, $\xi_r$, $\xi_R$ and $\lambda$ appear algebraically in $\delta h_+$, $\delta h_-$, $\delta h_v$ and $\delta a_R$, it is a simple exercise to build gauge invariant variables out of these quantities. In particular, we can introduce the following gauge invariant variables
\begin{subequations}
\begin{equation}
\hat{h}_{vv}(v,r)=h_{vv}(v,r)+\frac{k^2 f'(r) h_-(v,r)}{6 r}+\frac{f'(r) h_+(v,r)}{2 r}-2 \partial_vh_v(v,r)+\partial^2_v h_-(v,r)\,,
\end{equation}
\begin{multline}
\hat{h}_{vr}(v,r)=h_{vr}(v,r)-\frac{k^2 f'(r) h_-(v,r)}{3 r f(r)}-\frac{3 r f'(r)+k^2}{3 r f(r)}\partial_v h_-(v,r) -\frac{f'(r) h_+(v,r)}{r f(r)}
\\
+\frac{2 f'(r) h_v(v,r)}{f(r)}-\frac{\partial_v h_+(v,r)}{r f(r)}+\frac{2
   \partial_ vh_v(v,r)}{f(r)}-\frac{\partial_v^2h_-(v,r)}{f(r)}-2 \partial_r h_v(v,r)+\partial_v \partial_rh_-(v,r)\,,
\end{multline}
\begin{multline}
\hat{h}_{rr}(v,r)=h_{rr}(v,r)+\frac{k^2 \left[r f'(r)+f(r)\right]}{3 r^2 f(r)^2}h_-(v,r)+\frac{r f'(r)+f(r)}{r^2 f(r)^2}h_+(v,r)-\frac{2 f'(r) h_v(v,r)}{f(r)^2}
\\
+\frac{f'(r) \partial_v h_-(v,r)}{f(r)^2}-\frac{k^2
   \partial_r h_-(v,r)}{3 r f(r)}-\frac{\partial_r h_+(v,r)}{r f(r)}+\frac{2 \partial_r h_v(v,r)}{f(r)}-\frac{\partial_r \partial_v h_-(v,r)}{f(r)}\,,
\end{multline}
\begin{equation}
\hat{h}_r(v,r)=h_r(v,r)-\frac{k^2-6 f(r) }{6 r f(r)}h_-(v,r)-\frac{h_+(v,r)}{2 r f(r)}+\frac{h_v(v,r)}{f(r)}-\frac{\partial_v h_-(v,r)}{2 f(r)}-\frac{1}{2} \partial_r h_-(v,r)\,,
\end{equation}
\begin{equation}
\hat{a}_v(v,r)=a_v(v,r)-\partial_v a_R(v,r)+\frac{k^2 Q h_-(v,r)}{\sqrt{3} r^4}+\frac{\sqrt{3} Q h_+(v,r)}{r^4}\,,
\end{equation}
\begin{multline}
\hat{a}_r(v,r)=a_r(v,r)-\partial_r a_R(v,r)-\frac{k^2 Q h_-(v,r)}{\sqrt{3} r^4 f(r)}-\frac{\sqrt{3} Q h_+(v,r)}{r^4 f(r)}
\\
+\frac{2 \sqrt{3} Q h_v(v,r)}{r^3 f(r)}-\frac{\sqrt{3} Q \partial_v h_-(v,r)}{r^3 f(r)}\,.
\end{multline}
\label{eq:gaugein}%
\end{subequations}%
Inserting these definitions into the perturbed equations of motion yields a set of PDEs for the six gauge invariant variables $\hat{h}_{vv}$, $\hat{h}_{vr}$, $\hat{h}_{rr}$, $\hat{h}_{r}$, $\hat{a}_{v}$ and $\hat{a}_r$. These, in turn, we are able to solve in terms of two variables $\psi_{\pm}(v,r)$, via the following map
\begin{subequations}
\begin{multline}
\hat{h}_{vv}(v,r)=-\frac{18 r^5 f(r)^2}{L^4 \mathit{z}(r)}\partial_r \psi _-(v,r)-\frac{6 \sqrt{3} Q r^3 f(r)^2}{L^2
   \mathit{z}(r)} \partial_r \psi _+(v,r)\frac{18 f(r) L^2 r^5}{L^6 \mathit{z}(r)}\partial_v \psi _-(v,r)
   \\
   -\frac{6 \sqrt{3} Q f(r) L^2 r^4}{L^4 r
   \mathit{z}(r)} \frac{\partial \psi _+(v,r)}{\partial v}-\frac{9 r^2}{k^2 L^4}\partial_v^2\psi _-(v,r)+\frac{\mathit{z}_1(r)}{L^8 r^4 \mathit{z}(r)^2}\psi _-(v,r)-\frac{2 \sqrt{3} Q f(r)  \psi _+(v,r)}{L^4 \mathit{z}(r)^2}\mathit{z}_2(r)
\end{multline}
\begin{multline}
\hat{h}_{vr}(v,r)=\frac{\mathit{z}_6(r) \partial_v \psi _-(v,r)}{2 k^2 L^6 r^4 f(r) \mathit{z}(r)}+\frac{9 r^2 f(r) \partial_r^2\psi _-(v,r)}{2 k^2 L^4}+\frac{9 r^2 \partial^2_v\psi_-(v,r)}{k^2 L^4 f(r)}+\frac{\mathit{z}_3(r)
   \psi _-(v,r)}{2 L^8 r^4 f(r) \mathit{z}(r)^2}
   \\
   +\frac{12 \sqrt{3} Q r^3 f(r) \partial_r \psi _+(v,r)}{L^2 \mathit{z}(r)}+\frac{9 \mathit{z}_5(r) \partial_v\psi _-(v,r)}{2 k^2 L^6 r^3 \mathit{z}(r)}+\frac{2
   \sqrt{3} Q \mathit{z}_4(r) \psi _+(v,r)}{L^4 \mathit{z}(r)^2}+\frac{6 \sqrt{3} Q r^3 \partial_v \psi _+(v,r)}{L^2 \mathit{z}(r)}
\end{multline}
\begin{multline}
\hat{h}_{rr}(v,r)=-\frac{3 \mathit{z}_9(r)}{2 k^2 L^6 r^3 f(r) \mathit{z}(r)}\partial_r\psi _-(v,r)-\frac{9\left[L^2 \left(10 M r^2-9 Q^2\right)+3 r^6\right]}{2 k^2 L^6 r^3 f(r)^2}\partial_v \psi _-(v,r)
\\
-\frac{3
   \mathit{z}_7(r) \psi _-(v,r)}{2 L^8 r^4 f(r)^2 \mathit{z}(r)}+\frac{2 \sqrt{3} Q \mathit{z}_8(r) \psi _+(v,r)}{L^4 r^4 f(r) \mathit{z}(r)}-\frac{9 r^2 \partial_r^2\psi _-(v,r)}{2 k^2 L^4}-\frac{6 \sqrt{3}Q r^3 \partial_r \psi _+(v,r)}{L^2 \mathit{z}(r)}
\end{multline}
\begin{multline}
\hat{h}_r(v,r)=\frac{9 r^2}{2 k^2 L^4 f(r)} \partial_v\psi_-(v,r)+\frac{3 r \left[L^2 \left(k^2 r^4-3 Q^2\right)+6 r^6\right]}{2 L^6 f(r) \mathit{z}(r)}\psi _-(v,r)
\\
+\frac{9 r^2}{2 k^2 L^4} \partial_r\psi_-(v,r)+\frac{3
   \sqrt{3} Q r^3}{L^2 \mathit{z}(r)} \psi_+(v,r)
\end{multline}
\begin{multline}
   \hat{a}_v(v,r)=-\frac{2 r^2 f(r) \left(k^2 r^2+12 M\right)}{L^2 \mathit{z}(r)} \psi _+(v,r)-\frac{r f(r)}{L^2} \partial_r\psi_+(v,r)
   \\
   -\frac{3 \sqrt{3} Q \left[L^2 \left(k^2 r^4-3 Q^2\right)+6 r^6\right]}{L^6
   r^2 \mathit{z}(r)}\psi _-(v,r)-\frac{r}{L^2}\partial_v  \psi _+(v,r)
\end{multline}
\begin{multline}
\hat{a}_r(v,r)=\frac{9 \sqrt{3} Q }{k^2 L^4 r f(r)}\partial_v\psi _-(v,r)-\frac{3 \sqrt{3} Q  \left[L^2 \left(3 Q^2-k^2 r^4\right)-6 r^6\right]}{L^6 r^2 f(r) \mathit{z}(r)}\psi_-(v,r)
   \\
   +\frac{2}{L^2}\left[1+\frac{9
   Q^2}{\mathit{z}(r)}\right]\psi _+(v,r)+\frac{r}{L^2} \partial_r\psi_+(v,r)
\end{multline}
\label{eq:nuts}
\end{subequations}
so long as $\psi_{\pm}$ obey  the following coupled set of partial differential equations
\begin{subequations}
\begin{equation}
\begin{aligned}
&\Box\psi_+(v,r)=\frac{1}{r^2\mathit{z}(r)^2}\left[V_{++}(r)\psi_+(r)+V_{+-}(r)\psi_-(r)\right]
\\
&\Box\psi_-(v,r)=\frac{1}{r^2\mathit{z}(r)^2}\left[V_{-+}(r)\psi_+(r)+V_{--}(r)\psi_-(r)\right]
\end{aligned}
\end{equation}
where the polynomials $z_i(r)$ are listed in Appendix \ref{app:pol}, $\Box$ is calculated on an RN black hole in Eddington-Finkelstein coordinates and
\begin{equation}
\begin{aligned}
&V_{++}(r)=k^6 r^8+2 k^4 Q^2 r^4+48 M^2 \left(6 Q^2-k^2 r^4\right)+45 k^2 Q^4+16 M \left(k^4 r^6-6 k^2 Q^2r^2\right)
\\
&\qquad\qquad\qquad-\frac{4 r^4}{L^2} \left[k^4 r^6+9 r^2 \left(16 M^2-5 k^2 Q^2\right)+24 k^2 M r^4-324 M Q^2\right]-1152 M^3 r^2\,,
\\
&V_{+-}(r)=\frac{12 \sqrt{3} Q \left[3 k^2 r^8-L^2 \left(k^2 r^2+6 M\right) \left(12 M r^2-k^2 r^4-3 Q^2\right)-36 M r^6+81 Q^2 r^4\right]}{L^4}\,,
\\
&V_{-+}(r)=\frac{L^4k^2}{27}V_{+-}(r)\,,
\\
&V_{--}(r)=k^2 \left[k^4 r^8-\frac{4 \left(k^2 r^{10}+36 M r^8-45 Q^2 r^6\right)}{L^2}+2 Q^2 \left(k^2 r^4-72 M r^2\right)+48 M^2 r^4+45 Q^4\right]\,.
\end{aligned}
\end{equation}
\label{eqs:EFPhipm}
\end{subequations}
These equations reproduce the results of \cite{Jansen:2019wag} in the special case 
$n=3$. They can furthermore be recast into a set of \emph{decoupled} equations, thereby recovering the seminal construction of Kodama and Ishibashi \cite{Kodama:2003kk}. This refinement will not play a central role in what follows, but for completeness we present the relation between our variables and the standard Kodama–Ishibashi ones:
\begin{subequations}
\begin{equation}
\begin{aligned}
&\psi_+(v,r)=\hat{\psi}_+(v,r)-3 \sqrt{3} \eta\frac{\hat{\psi}_-(v,r)}{L^2}
\\
&\psi_-(v,r)=\frac{k^2}{3 \sqrt{3}} \eta  L^2\hat{\psi}_-(v,r)+\hat{\psi}_+(v,r)
\end{aligned}
\end{equation}
with
\begin{equation}
\eta\equiv\frac{Q}{2 M+\sqrt{k^2 Q^2+4 M^2}}\,.
\end{equation}
\end{subequations}%
where $\hat{\psi}_{\pm}(v,r)$ are the Kodama–Ishibashi variables introduced in \cite{Kodama:2003kk}.

 After decoupling, we obtain equations of the form \eqref{eq:wavelike} for $\hat{\psi}_\pm(v,r)$ with potentials $\hat{V}_\pm(r)$ which are given by eigenvalues of a symmetric matrix $(V_{\pm \pm})$. As discussed in Sec. \ref{sec:toy_model}, the late-time behavior at the extremal horizon is governed by the value of the potential at $r_+$. These values are given by 
\begin{equation}
    \hat{V}_\pm(r_+) \equiv\mu^2_{\pm} = \frac{1}{r_+^2}\left(k^2+12\frac{r_+^2}{L^2} \pm \frac{r_+}{L} \sqrt{32 k^2 + 144 \frac{r_+^2}{L^2}}\right).
\end{equation}
$\mu_+$ is a growing function of $k$ with minimum at $k=0$:
\begin{equation}
    \mu_+^2 (k=0) = \frac{24}{L^2}.
\end{equation}
Using \eqref{eq:alpha} this corresponds to $\alpha = 2$ and so $+$ modes are quickly decaying along the horizon. $-$ modes are more interesting.   These modes have scaling exponents:
\begin{equation}\label{eq:EMmodes}
    \alpha(k) =\frac{1}{2} \left(1+\sqrt{\frac{k^2 L^2-4  r_+ \sqrt{2 k^2 L^2+{9 r_+^2 }}}{3r_+^2}+5}\right)
\end{equation}
In particular,
\begin{equation}
    \mu_-^2(k=0) = 0,
\end{equation}
which corresponds to $\alpha = 1$. However, this is not a minimum of $\mu_-$ (and so, not a minimum of $\alpha(k)$). Instead, we have
\begin{subequations}
    \begin{equation}
        \min \mu_-^2 = -\frac{1}{2L^2}< 0
    \end{equation}
    at
    \begin{equation}
        k_{\min} = \frac{r_+}{L} \sqrt{\frac{7}{2}},
    \end{equation}
    which corresponds to
    \begin{equation}
        \alpha_{\min} = \frac{1}{2} \left(1 + \sqrt\frac{5}{6} \right) \approx 0.956435< 1.
    \end{equation}
\end{subequations}%
It follows that the master scalar for the $-$ scalar sector decays slower than $1/v$. In particular, its first derivative will blow up along the horizon. 

 It is perhaps surprising that a particular nonzero $k$ is selected by the dynamics when the background is translation invariant. However this type of breaking of translation invariance has been seen before. When charged scalar fields are present outside a nonextreme black hole, the first modes to become unstable as one approaches extremality can have nonzero $k$ \cite{Iqbal:2011ae}.  This particular value of $k$ can be also understood as follows. If we dimensionally reduce the near-horizon region to $\textrm{AdS}_2$, different $k$ components would correspond to scalar fields with masses squared $\mu_{\pm}^2(k)$. As explained earlier, $\alpha(k)$ is the same as the conformal dimension of these perturbations. We thus learn that in dimensions five and higher, certain perturbations are relevant (they satisfy $\alpha <1$) and $k_{\min}$ corresponds to the most relevant one.\footnote{The fact that higher-dimensional spherical extremal Reissner--Nordstr\"om black holes admit relevant deformations (and are thus RG unstable) was first observed in \cite{Horowitz:2022leb}.}

In order to understand the boundary conditions, we need to determine the asymptotic behaviour of the $\psi_{\pm}(v,r)$ near the conformal boundary, i.e. near $r=+\infty$. A Frobenius analysis reveals that
\begin{equation}
\psi_{\pm}(v,r)=\sum_{i=2}^{+\infty}\frac{\psi_{\pm}^{(i)}(v)}{r^i}+\log r\sum_{i=2}^{+\infty}\frac{\tilde{\psi}_{\pm}^{(i)}(v)}{r^i}\,
\end{equation}
with $\psi_{\pm}^{(i)}(v)$ and $\tilde{\psi}_{\pm}^{(i)}(v)$ for $i>2$ fixed in terms of $\psi^{(2)}_{\pm}(v)$, $\tilde{\psi}^{(2)}_{\pm}(v)$ and its derivatives with respect to $v$. For instance, for $i=3$ we find
\begin{equation}
\psi^{(3)}_{\pm}(v)=L^2\partial_v\psi^{(2)}_{\pm}(v)\quad\text{and}\quad \tilde{\psi}^{(3)}_{\pm}(v)=L^2\partial_v\tilde{\psi}^{(2)}_{\pm}(v)\,.
\end{equation}

We restrict attention to physical situations in which the boundary metric remains flat and no external boundary electric field is applied. A straightforward exercise shows that this requirement forces $\tilde{\psi}^{(2)}_{\pm}(v)=0$. In turn, this condition eliminates any $\log r$ in the large $r$ expansion of $\psi_{\pm}(v,r)$. What remains, then, is to interpret the physical significance of $\psi^{(2)}_{\pm}(v)$. This is best understood by calculating the holographic stress energy tensor as a function of $\psi^{(2)}_{\pm}(v)$ and derivatives with respect to $v$.

We carry out this procedure in two steps. First, we transform the system into Fefferman–Graham coordinates. Second, following \cite{deHaro:2000vlm}, we extract the corresponding holographic stress–energy tensor and holographic current. To change the background metric to Fefferman–Graham coordinates up to order $z^2$, we take
\begin{equation}
R=\frac{\hat{R}}{L}\,,\quad v=t-z-\frac{3 M z^5}{20 L^6}\quad\text{and}\quad r=\frac{L^2}{z} \left(1+\frac{M z^4}{4 L^6}\right)\,
\end{equation}
and introduce $\hat{k}\equiv k L$. For the metric perturbation, we follow a slightly different—but equivalent—approach. We impose the conditions $h_{za} = 0$ and $a_z = 0$. From Eq.~\eqref{eq:nuts}, we obtain a mapping between the gauge-invariant quantities and $\psi_{\pm}(v,r)$. By inverting Eq.~(\ref{eq:gaugein}), we then express $h_{vv}$, $h_{vr}$, $h_{rr}$, $h_r$, $a_v$, and $a_r$ as functions of $\psi_{\pm}(v,r)$, $h_{\pm}$, $h_v$, and $a_R$. We treat $h_{\pm}, h_v$, and $a_R$ as parametrising the gauge redundancy, encompassing both diffeomorphisms and $U(1)$ transformations. To transform to Fefferman-Graham coordinates, we apply the background coordinate transformation to the metric perturbation, changing from $(v,r)$ to $(t,z)$ coordinates, and then set
\begin{equation}
h_{\pm}(v,r)=\sum_{i=0}^{+\infty}\frac{h_{\pm}^{(i)}(v)}{r^i}\,,\quad h_v(v,r)=\sum_{i=0}^{+\infty}\frac{h_{v}^{(i)}(v)}{r^i}\quad\text{and}\quad a_R(v,r)=\sum_{i=0}^{+\infty}\frac{a_{R}^{(i)}(v)}{r^i}\,,
\end{equation}
and adjust the various coefficients order by order to ensure that $h_{za} = 0$ and $a_z = 0$. Once this is achieved, we follow \cite{deHaro:2000vlm} to extract the corresponding components of the holographic stress-energy tensor and current. After following this procedure we find
\begin{multline}
\langle T_{tt}\rangle=\frac{3 M}{2 L^6}-\frac{3 \tilde{p}^2}{4 L^8} \psi_-^{(2)}(t)\mathbb{Y}_{\hat{k}}\,,\quad \langle T_{t\hat{R}}\rangle=\frac{3}{4 L^{10}}\psi_-^{(3)}(t) \Grad_{\hat{R}}\mathbb{Y}_{\hat{k}}\,,
\\
\langle J^t\rangle=\frac{\sqrt{3} Q}{L^4}+\frac{\tilde{p}^2}{2 L^4}\psi_+^{(2)}(t)\mathbb{Y}_{\hat{k}}(\hat{R})\,,\quad\text{and}\quad \langle J^{\hat{R}}\rangle=-\frac{1}{2 L^6}\psi_+^{(3)}(t)\Grad_{\hat{R}}\mathbb{Y}_{\hat{k}}\,.
\label{eq:match}
\end{multline}
All remaining components of the holographic stress-energy tensor can likewise be determined, but they are not relevant for our current analysis. The above expressions clarify the significance of $\psi_{\pm}^{(2)}(v)$ and $\psi_{\pm}^{(3)}(v)$, which correspond to deformations in the energy density, fluid velocity along $\hat{R}$, and the charge density and current along $\hat{R}$.

The solution of Eqs.~(\ref{eqs:EFPhipm}) requires a precise description of how Eddington-Finkelstein coordinates are employed to implement the time integration along the $v$ direction. To this end, we introduce the directional derivatives
\begin{equation}
{\rm d}_v\psi_{\pm}=\partial_v\psi_{\pm}+\frac{f(r)}{2}\,\partial_r\psi_{\pm}\,,
\end{equation}
in terms of which the equations for $\psi_{\pm}$ take the form
\begin{equation}
\begin{aligned}
&\partial_r {\rm d}_v\psi_{+}+\frac{3}{2r}{\rm d}_v\psi_{+}
=\frac{1}{2r^2 z(r)^2}\Big[V_{++}(r)\psi_{+}(v,r)+V_{+-}(r)\psi_{-}(v,r)\Big]
-\frac{3f(r)}{4r}\,\partial_{r}\psi_{+}\,, \\[4pt]
&\partial_r {\rm d}_v\psi_{-}+\frac{3}{2r}{\rm d}_v\psi_{-}
=\frac{1}{2r^2 z(r)^2}\Big[V_{-+}(r)\psi_{+}(v,r)+V_{--}(r)\psi_{-}(v,r)\Big]
-\frac{3f(r)}{4r}\,\partial_{r}\psi_{-}\,.
\end{aligned}
\end{equation}

The right-hand side of each equation depends only on $\psi_{\pm}(v,r)$. Hence, once $\psi_{\pm}(v,r)$ is specified at a given time, the equations can be integrated, provided that appropriate boundary conditions are imposed for ${\rm d}_v \psi_{\pm}$ at the conformal boundary. Having fixed the boundary conditions for $\psi_{\pm}(v,r)$, we can then infer the corresponding behavior of ${\rm d}_v \psi_{\pm}(v,r)$. From the asymptotic expansion above, one finds
\begin{equation}
{\rm d}_v\psi_{\pm}(v,r)=-\frac{\psi^{(2)}_{\pm}(v)}{L^2 r}+\mathcal{O}(r^{-2})\,.
\label{eq:bcs}
\end{equation}

Once ${\rm d}_v\psi_{\pm}(v,r)$ is known on a given time slice, the derivatives $\partial_v \psi_{\pm}(v,r)$ can be directly extracted from ${\rm d}_v \psi_{\pm}$. The system may then be evolved forward using any suitable time-integration scheme; in this work, we consistently employ an explicit fourth-order Runge-Kutta method.

To proceed, recall that the black hole radius $r_+$ is defined by 
\begin{equation}
f(r_+)=0 \;\;\Rightarrow\;\; M=\frac{r_+^2}{L^2}\,(1+\tilde{Q}^2)\,,
\end{equation}
where $\tilde{Q}\equiv L Q/r_+^3$, with extremality at $\tilde{Q}^2=2$. We then perform the coordinate redefinitions
\begin{equation}
r=\frac{r_+}{y}\,, \qquad v=\frac{L^2}{r_+}\,\hat{v}\,,
\label{eq:compact}
\end{equation}
and introduce the rescaled fields
\begin{equation}\label{eq:Qdef}
\begin{aligned}
\psi_+(v,r) &= \frac{L^3}{r_+}\left(\frac{r_+}{r}\right)^2 
Q_1\!\left(\frac{r_+ v}{L^2}, \frac{r_+}{r}\right), 
&\quad \psi_-(v,r) &= L^4 \left(\frac{r_+}{r}\right)^2 
Q_2\!\left(\frac{r_+ v}{L^2}, \frac{r_+}{r}\right), \\[4pt]
{\rm d}_v\psi_+(v,r) &= L\left(\frac{r_+}{r}\right) 
Q_3\!\left(\frac{r_+ v}{L^2}, \frac{r_+}{r}\right), 
&\quad {\rm d}_v\psi_-(v,r) &= r_+ L^2\left(\frac{r_+}{r}\right) 
Q_4\!\left(\frac{r_+ v}{L^2}, \frac{r_+}{r}\right).
\end{aligned}
\end{equation}

The resulting equations for $Q_I$ ($I=1,2,3,4$) depend only on the variables 
$\hat{v}$ $y$, $\tilde{Q}$ and $\tilde{k}\equiv Lk/r_+$. In these coordinates, the conformal boundary is located at $y=0$, while the horizon lies at $y=1$. To determine $Q_3$ and $Q_4$ on a given time slice, we require boundary conditions. These stem from Eq.~(\ref{eq:bcs}) and turn out to be simply
\begin{equation}
Q_3(\hat{v},0)=Q_1(\hat{v},0)\quad \text{and}\quad Q_4(\hat{v},0)=Q_2(\hat{v},0)\,.
\end{equation}
The full integration can then be carried out once $Q_1(\hat{v},y)$ and $Q_2(\hat{v},y)$ are specified on a given time slice.

Finally, we now show that the relevant components of the curvature which characterize the tidal forces experienced by an observer falling into the perturbed extremal horizon, can be expressed in terms of $\psi_{\pm}$.  We first note that in the perturbed geometry, the following vector is tangent to future-directed affinely parametrised null geodesics
\begin{subequations}
\begin{equation}
U=\frac{1}{2} h_{rr}(v,r)\mathbb{Y}_k(R) \frac{\partial}{\partial v}-\left[1-\lambda_r(v,r) \mathbb{Y}_k(R)\right]\frac{\partial}{\partial r}+\lambda_R(v,r) \mathbb{S}^k_I\frac{\partial}{\partial x^I}
\end{equation}
with
\begin{equation}
h_{r}(v,r)-2 \partial_rh_r(v,r)+2 r^2 \partial_r\lambda_R(v,r)+4 r \lambda _R(v,r)=0\,,
\end{equation}
\begin{equation}
f'(r) h_{rr}(v,r)+f(r) \partial_rh_{rr}(v,r)-\partial_vh_{rr}(v,r)+\partial_rh_{vr}(v,r)-2 \partial_r\lambda_r(v,r)=0\,.
\end{equation}
\end{subequations}

Once we have $U$, we can define a tidal tensor $\widehat B$, with components
\begin{equation}
\widehat{B}_{IJ}\equiv U^aU^bR_{aI bJ}-\frac{g_{I J}}{3}U^a U^bg^{PQ}R_{aPbQ}\,,
\end{equation}
for which
\begin{equation}
\widehat{B}^{I}_{\phantom{I}J}=\mathrm{diag}\{-2\Phi,\Phi,\Phi\}\,,
\label{eq:lineartidal}
\end{equation}
where $R_{abcd}$ are the components of the Riemann tensor. Here, $I$ and $J$ range over $R$ and the $S^2$ directions, i.e., the spatial directions on the brane. It is worth noting that $\widehat{B}_{IJ}$ can equivalently be defined via the Weyl tensor. One may ask why we focus on the components of the traceless part of the Riemann tensor rather than the full tensor. There are three main reasons: (i) this quantity is gauge invariant and therefore depends only on the $\psi_{\pm}$, (ii) through its definition in terms of the Weyl tensor, it naturally connects to the generalisation of the Teukolsky scalars to higher dimensions \cite{Durkee:2010xq} and (iii) the pure trace component turns out to vanish in the linear approximation.

Using our expressions for the several metric components in terms of $\psi_{\pm}$, we find 
\begin{subequations}
\begin{equation}
\Phi(v,r)=\Phi_k(v,r)\left[\Grad_R\Grad_R\mathbb{Y}_k+\frac{k^2}{3}\mathbb{Y}_k\right]\,,
\label{eq:linear}
\end{equation}
with
\begin{multline}
\Phi_k(v,r)=\frac{1}{2z(r) L^2}\Bigg[-9 \sqrt{3} Q \psi _+(v,r)-6 \sqrt{3} Q r \partial_r\psi_+(v,r)-\frac{27 r^2}{L^2} \psi _-(v,r)
\\
-\frac{27}{L^2 k^2 r} \left(k^2 r^4+4 M r^2-3Q^2\right)\partial_r\psi_-{}(v,r)-\frac{9}{2 L^2 k^2}z(r)\partial^2_r\psi_-(v,r)\Bigg]\,.
\end{multline}
\end{subequations}%

\subsubsection{Working in linear Bondi-Sachs coordinates to match nonlinear code}
When proceeding to the nonlinear calculation we will use Bondi-Sachs coordinates. These are (partially) defined via $h_{r}(v,r)=0$, $h_{rr}(v,r)=0$ and fixing the \emph{radial} dependence of $h_+(v,r)$. Since we have a map between the gauge invariant variables, the potentials $\Phi_{\pm}$ and the several metric components, the two conditions $h_{r}(v,r)=0$ and $h_{rr}(v,r)=0$ give non-trivial relations among the $\Phi_{\pm}$, $h_{\pm}$ and $h_{v}$. From the condition $h_{r}(v,r)=0$ we can determine $h_v$ algebraically in terms of $\Phi_{\pm}$, $h_{\pm}$ and their derivatives. From the condition $h_{rr}=0$, we find
\begin{subequations}
\begin{multline}
\frac{2 \partial_rh_-(v,r)}{r}-\frac{9 r}{L^4 k^2}\left[\frac{6 k^2 r^3 \psi_-(v,r)+6 \left(4 M r^2+k^2 r^4-3 Q^2\right) \partial_r\psi _-(v,r)}{z(r)}+r \partial^2_r\psi_-(v,r)\right]
\\
-\frac{2 h_-(v,r)}{r^2}-\frac{6 \sqrt{3} Q r^2}{L^2 z(r)}\left[3 \psi _+(v,r)+2 r \partial_r\psi_+(v,r)\right]-\partial^2_rh_-(v,r)=0
\end{multline}
This equation involves no time derivatives. Knowing $h_-$ at a particular moment allows us to relate $\psi_+$ and $\psi_-$ up to two integration constants. Specifically, $h_-(0,r)$ is precisely one of the initial bulk data in the Bondi-Sachs evolution scheme. The other variable that we specify at a give moment of time is $F_{rR}$, which we can readily compute in terms of $h_{\pm}(v,r)$ and $\psi_{\pm}$. Since $F_{rR}$ transforms as a vector on the three sphere, we expand $F_{rR}(v,r,R)=\hat{F}_{rR}(v,r)\Grad_{R}\mathbb{Y}_k$, and find
\begin{equation}
F_{rR}(v,r)=-\frac{2 \sqrt{3} Q h_-(v,r)}{r^4}-\frac{2 \psi _+(v,r)}{L^2}+\frac{\sqrt{3} Q \partial_rh_-(v,r)}{r^3}-\frac{r \partial_r\psi_+(v,r)}{L^2}+\frac{9 \sqrt{3} Q \partial_r\psi_-(v,r)}{L^4 k^2 r}\,.
\end{equation}
\label{eq:nullmatch}
\end{subequations}
Again, the expression above involves only $r$-derivatives. The two equations (\ref{eq:nullmatch}) allow us to determine $\psi_{\pm}(v,r)$ from $\hat{F}_{rR}(v,r)$ and $h_-(v,r)$ at a given instant, up to three integration constants. For a given $\hat{F}_{rR}(v,r)$ profile, $\psi_+^{(3)}(v)$ is fixed, but $\psi_{\pm}^{(2)}(v)$ and $\psi_{-}^{(3)}(v)$ are not. These require specifying the energy density, the fluid velocity along $R$, and the charge density at that $v$ (see Eq.~(\ref{eq:match})). In this work, we will focus on initial deformations for which the fluid velocity along $R$ and the change in the energy density and charge density vanish initially. 
\section{Non-linear evolution\label{sec:nonl}}
We consider five-dimensional spacetimes with an $SO(3)$ isometry group whose generic orbits are two-dimensional, compact, and simply connected. These orbits are necessarily round two-spheres, allowing the metric to be expressed in coordinates adapted to them, with line element denoted by ${\rm d}\Omega_2^2$.

We work in Bondi-Sachs coordinates, in which our line element takes the following form
\begin{subequations}\label{eq:nonlinear}
\begin{equation}
{\rm ds}^2=\frac{L^2}{y^2}\left[-e^{2\beta}\left(V{\rm d}v^2+2{\rm d}v{\rm d}y\right)+e^{2\chi}h_{IJ}({\rm d}x^I-U^I{\rm d}v)({\rm d}x^J-U^J{\rm d}v)\right]\,.
\end{equation}
For our particular setup, we take
\begin{equation}
h_{IJ}({\rm d}x^I-U^I{\rm d}v)({\rm d}x^J-U^J{\rm d}v)\equiv \frac{1}{4xA^2}\left({\rm d}x-U^x\,{\rm d}v\right)^2+x A{\rm d}\Omega_2^2\,.
\end{equation}
\end{subequations}%
Note that $x$ is a radial coordinate on the boundary and $y$ is a radial coordinate in the bulk.
In the equations above, $A$, $V$, $U^x$ $\chi$ and $\beta$ are functions of $(v,x,y)$ to be determined by our integration scheme, and upper case Latin indices run over the spatial boundary directions.

With our metric ansatz, the system remains underdetermined. By an appropriate redefinition of the radial coordinate $y$, we can eliminate either $\chi$ or $\beta$. Writing $\chi$ in the form
\begin{equation}
\chi(v,y,x)=-\log[1+y^4\chi_4(v,x)]
\label{eq:pg}
\end{equation}
we can use the residual freedom  in $\chi_4$ to ensure that the apparent horizon is located at $y=1$. This particular choice of $\chi$ is loosely motivated by the fact that black brane solutions in $D=5$ can be expressed in a form where $\chi$ takes the structure above.

Assuming the apparent horizon can indeed be fixed at $y=1$, the integration domain is restricted to $y\in[0,1]$ and $x\in \mathbb{R}^+$, with the conformal boundary at $y=0$ and the apparent horizon at $y=1$. The axis $x=0$ corresponds to the point where the round two-sphere shrinks to zero size, while $x=+\infty$ represents the asymptotic region on the conformal boundary. Finally, we note that $\chi$ controls the volume of constant-$v$ slices: since $h$ has fixed determinant, our gauge choice for $\chi$ already partially fixes this volume factor.

For the Maxwell field strength we take
\begin{equation}
F=L\left(F_{vy}{\rm d}v\wedge{\rm d}y+F_{vx}{\rm d}v\wedge{\rm d}x+F_{yx}{\rm d}y\wedge{\rm d}x\right)\,.
\end{equation}
To guarantee, via the Poincar\'e lemma, that a one-form potential $A$ exists so that $F={\rm d}A$, we will impose the Bianchi identities in our integration scheme. For our symmetry class, this reads
\begin{equation}
\partial_yF_{vx}-\partial_x F_{vy}-\partial_v F_{yx}=0\,.
\label{eq:bianchi}
\end{equation}
\subsection{The apparent horizon}
In Bondi–Sachs coordinates, $\partial/\partial y$ is globally null. This property will be particularly useful when computing tidal forces on the apparent horizon, which we will require to be located at $y=1$. To this end, we compute the expansion of a congruence of affinely parametrised null geodesics with tangent vector $\xi^a$, writing
\begin{equation}
\xi_a\xi^a=0\quad\text{and}\quad \xi^a\nabla_a\xi_b=0\,.
\end{equation}
The expansion $\Theta$ is simply given by $\Theta=\nabla_a\xi^a$, with the apparent horizon being $\Theta=0$. We parametrise $\xi_a$ as the gradient of a hypersurface $S=0$, which is orthogonal to the geodesic congruence, together with an arbitrary rescaling function $f$:
\begin{equation}
\xi_a=f \nabla_aS\,.
\end{equation}
Even without specifying the explicit form of $S$, we can determine $\nabla_vS$ using the null condition $\xi_a \xi^a = 0$. Since this condition also requires $\nabla_aS \nabla^aS = 0$, the hypersurface $S=0$ is itself null. Furthermore, the geodesic condition $\xi^a \nabla_a\xi_b = 0$ implies orthogonality, $\nabla_a S \nabla^a f = 0$, which in turn allows us to solve for $\nabla_v f$. Using the aforementioned conditions, it is a simple exercise to compute $\Theta$ for $S=y-1$, which yields
\begin{equation}
\left.\Theta\right|_{y=1}=\left.\frac{3 e^{-2 \beta } f}{L^2}\left[\frac{1}{2}y^{-1}(y \partial_y \chi-1)V-\frac{1}{3}D_IU^I-\partial_v \chi\right]\right|_{y=1}\,,
\label{eq:appa}
\end{equation}
where $D$ denotes the Levi-Civita connection compatible with the metric $h_{IJ}$.
\subsection{Asymptotic expansion near the conformal boundary}
Let us define the following tensors
\begin{equation}
E_{ab}\equiv R_{ab}+\frac{4}{L^2}g_{ab}-\frac{1}{2}\left(F_{ac}F_{b}^{\phantom{b}c}-\frac{1}{6}g_{ab}F_{cd}F^{cd}\right)\quad \text{and}\quad J^a=\nabla_bF^{ba}\,,
\end{equation}
so that the trace-reversed Einstein and Maxwell equations read
\begin{equation}
E_{ab}=0\quad\text{and}\quad J^a=0\,,
\end{equation}
respectively.

In what follows it is useful to define the momentum $\Pi_I$ conjugate to $U^I$ as
\begin{subequations}
\begin{equation}
\Pi_I\equiv \frac{e^{3\chi-2\beta}}{y^3}h_{IJ}\partial_y U^J
\label{eq:momPi}
\end{equation}
as well as a momentum associated with the Maxwell field
\begin{equation}
\Pi_F\equiv \frac{e^{3\chi-2\beta}}{y}(F_{vy}-F_{yx}U^x)\,.
\label{eq:momPF}
\end{equation}
\end{subequations}%
We note that only the $x$ component of $\Pi^I$ is nonzero for our setup.

In addition, we also define the following auxiliary variables
\begin{equation}
\begin{aligned}
&{\rm d}_{v}\chi\equiv \partial_v \chi-\frac{V}{2y}(y\partial_y \chi-1)\,,
\\
&{\rm d}_{v}A\equiv\partial_v A-\frac{V}{2}\partial_y A\,,
\\
&{\rm d}_{v}F_{vx}\equiv F_{vx}-\frac{V}{2}F_{yx}\,,
\end{aligned}
\end{equation}
Substituting ${\rm d}_{v}\chi$ into Eq.~(\ref{eq:appa}), the apparent horizon boundary condition $\left.\Theta\right|_{y=1} = 0$ becomes
\begin{equation}
\left.{\rm d}_v\chi+\frac{1}{3}D_IU^I\right|_{y=1}=0\,.
\label{eq:appa1}
\end{equation}

To proceed, we need to detail the asymptotic expansion near the conformal boundary located at $y=0$. We do not want to turn on any source at the conformal boundary, which makes our expansion considerably simple\footnote{If a source is turned on at the conformal boundary - for example, a boundary electric field or a boundary metric deviating from exact Minkowski - the expansion acquires logarithmic terms, which present significant challenges for certain numerical methods such as global collocation methods.}. The expansion of the several metric and gauge field variables reads
\begin{subequations}
\begin{equation}
\begin{aligned}
A & = 1+A_4(v,x) y^4+\left[\frac{2 U_4(v,x)}{15 x}-\frac{2}{15} \partial_x U_4(v,x)+\partial_v A_4(v,x)\right] y^5+\mathcal{O}(y^6)\,,
\\
\beta&=1+\frac{3}{2} \chi _4(v,x) y^4+\mathcal{O}(y^6)\,,
\\
U^x&=U_4(v,x) y^4+\left[\frac{48}{5} A_4(v,x)+\frac{32}{5} x \partial_x A_4(v,x)+4 x \partial_x \chi _4(v,x)\right] y^5+\mathcal{O}(y^6)\,,
\\
V&=1+V_4(v,x) y^4+\left[\frac{U_4(v,x)}{3 x}+\frac{2}{3} \partial_x U_4(v,x)+2 \partial_v \chi _4(v,x)\right] y^5+\mathcal{O}(y^6)\,,
\\
F_{vy}&=-2 g_2(v,x) y-2 \partial_v g_2(v,x) y^2+\mathcal{O}(y^3)\,,
\\
F_{yx}&=2 p_2(v,x) y+\left[3 \partial_v p_2(v,x)-\partial_x g_2(v,x)\right] y^2+\mathcal{O}(y^3)\,,
\\
F_{vx}&=\left[\partial_v p_2(v,x)-\partial_x g_2(v,x)\right] y^2+\mathcal{O}(y^3)\,,
\end{aligned}
\end{equation}
together with
\begin{equation}
\begin{aligned}
&\partial_v V_4(v,x)-\frac{2 U_4(v,x)}{3 x}-\frac{4}{3} \partial_x U_4(v,x)+5 \partial_v \chi _4(v,x)=0\,,
\\
&\partial_v U_4(v,x)-12 A_4(v,x)-8 x \partial_x A_4(v,x)-x \partial_x V_4(v,x)-5 x \partial_x \chi _4(v,x)=0\,,
\\
&\partial_v g_2(v,x)-6 p_2(v,x)-4 x \partial_x p_2(v,x)=0\,.
\end{aligned}
\label{eqs:cons}
\end{equation}
\end{subequations}%
The first two equations in Eqs.~(\ref{eqs:cons}) correspond to conservation of the holographic stress-energy tensor - respectively, the energy and momentum in the $x$ direction - while the last equation expresses charge conservation. To make this precise, we can compute the holographic stress energy tensor and one points functions for the current according to \cite{deHaro:2000vlm}. The first step is to change to Fefferman–Graham coordinates, in which the metric takes the simple form
\begin{equation}
{\rm d}s^2=\frac{L^2}{z^2}\left[{\rm d}z^2+{\rm d}s_0^2+z^2{\rm d}s_2^2+z^4{\rm d}s_4^2+\mathcal{O}(z^5)\right]\,,
\end{equation}
where ${\rm d}s_0^2$ is interpreted as the boundary metric, and ${\rm d}s_2^2$ is completely fixed in terms of ${\rm d}s_0^2$. Since our boundary metric is flat, ${\rm d}s_2^2$ turns out to vanish \cite{deHaro:2000vlm}. To change to this coordinates, up to order $z^4$, we set
\begin{equation}
\begin{aligned}
&v=t-z+\frac{3}{40} \left[V_4(t,X)+5 \chi _4(t,X)\right] z^5+\mathcal{O}(z^6)\,,
\\
&y=z+\frac{1}{8} \left[V_4(t,X)-3 \chi _4(t,X)\right] z^5+\mathcal{O}(z^6)\,,
\\
&x=X-\frac{1}{5} U_4(t,X)z^5 +\mathcal{O}(z^6)\,.
\end{aligned}
\end{equation}
Higher-order terms in the expansion can be adjusted so that $g_{z\mu}=0$ (with $\mu\in\{t,X,\Omega_i\}$) and $g_{zz}=L^2/z^2$ hold to higher orders around $z=0$.

In these coordinates, the boundary metric takes the form
\begin{equation}
{\rm d}s^2_0 = -{\rm d}t^2 + \frac{{\rm d}X^2}{4X} + X\,{\rm d}\Omega_2^2\,,
\end{equation}
which can be recognized as flat space upon identifying $X = R^2$. This substitution brings the metric to the standard canonical form of $\mathbb{M}^{1,3}$ in spherical polar coordinates:
\begin{equation}
{\rm d}s^2_0 = -{\rm d}t^2 + {\rm d}R^2 + R^2\,{\rm d}\Omega_2^2\,.
\end{equation}

A similar expansion holds for the Maxwell field strength, whose non-zero components to linear order in $z$ are
\begin{equation}
\begin{aligned}
&F_{tz}=-2 L g_2(t,X)z+\mathcal{O}(z^2)\,,
\\
&F_{Xz}=-2 L p_2(t,X)z+\mathcal{O}(z^2)\,.
\end{aligned}
\end{equation}

From the above expansion in Fefferman–Graham for both the metric and the gauge field, it is a simple exercise to compute both the expectation value of the holographic stress energy tensor and the expectation values for the current
\begin{subequations}
\begin{equation}
\begin{aligned}
&\langle T_{tt}\rangle=-\frac{3}{4} \left[V_4(t,X)+5 \chi _4(t,X)\right]
\\
&\langle T_{tX}\rangle=-\frac{U_4(t,X)}{4 X}
\\
&\langle T_{XX}\rangle=-\frac{8 A_4(t,X)+V_4(t,X)+5 \chi _4(t,X)}{16 X}
\\
&\langle T_{\Omega_i\Omega j}\rangle\equiv\langle T_{\Omega}\rangle G_{\Omega_i\Omega_j}=X\left[A_4(t,X)-\frac{1}{4} V_4(t,X)-\frac{5}{4} \chi _4(t,X)\right]G_{\Omega_i\Omega_j}
\end{aligned}
\end{equation}
where $\Omega_i$ parameterize the round $S^2$, and $G_{\Omega_i\Omega_j}$ denotes the canonical metric of the unit-radius sphere and
\begin{equation}
\begin{aligned}
&\langle J^t\rangle=-g_2(t,X)
\\
&\langle J^x\rangle=-p_2(t,X)
\\
&\langle J^{\Omega_i}\rangle=0
\end{aligned}
\end{equation}
\end{subequations}%
From the above, it is clear that $g_2(t,X)$ can be interpreted as a charge density, while $p_2(t,X)$ represents a current along the $X$ direction. $V_4$, up to factors of $\chi_4$, is related to the energy density of the field theory, whereas $A_4$ parametrizes a form of anisotropic pressure on the $S^3$. It is a simple exercise to show that the conservation of the holographic stress energy tensor and current follow directly from Eqs.~(\ref{eqs:cons}).

For a RN black hole in five spacetime dimensions, we simply have
\begin{equation}
\begin{aligned}
&V(v,y,x)=\left(1-y^2\right) \left(1+y^2-\frac{\lambda _0^2 y^4}{12} \right)\,,\quad F_{vy}=\lambda_0 y\,,\quad A(v,y,x)=1\,,
\\
& U^x(v,y,x)=\chi_4(v,x)=\beta(v,y,x)=F_{yx}(v,y,x)=F_{vx}(v,y,x)=0\,,
\end{aligned}
\end{equation}
yielding
\begin{equation}
V_4=-1-\frac{\lambda _0^2}{12}\,,\quad g_2=-\frac{\lambda _0}{2}\quad\text{and}\quad p_2=0\,.
\end{equation}
From $V$, we compute the temperature of the black hole with respect to $\partial/\partial v$
\begin{equation}
T=\frac{24-\lambda _0^2}{24 \pi }
\end{equation}
and we find that extremality requires $|\lambda_0|=2\sqrt{6}$. In what follows, we will use $\lambda_0=-2\sqrt{6}$.
\subsection{Marching orders and numerical variables}
For numerical convenience, it is useful to introduce $B \equiv e^{\beta}$, along with the following tilded variables
\begin{equation}
\begin{aligned}
A(v,x,y)&\equiv1+y^4 x\tilde{A}(v,x,y)\,,
\\
B(v,x,y)&\equiv1+y^4\left[\frac{3}{2}\chi_4(v,x)+y\tilde{B}(v,x,y)\right]\,,
\\
F_{yx}(v,x,y)&\equiv y \tilde{F}_{yx}(v,x,y)\,,
\\
\Pi_F(v,x,y)&\equiv 2\sqrt{6}+y \tilde{\Pi}_F(v,x,y)\,,
\\
\Pi^x(v,x,y)&\equiv\tilde{\Pi}^x(v,x,y)\,,
\\
U^x(v,x,y)&\equiv y^4 x\tilde{U}^x(v,x,y)\,,
\\
F_{vy}(v,x,y)&\equiv 2\sqrt{6}y+y\tilde{F}_{vy}(v,x,y)\,,
\\
{\rm d}_v\chi(v,x,y)&\equiv\frac{1}{2 y}-\frac{3 y^3}{2}+y^5+y^3\widetilde{{\rm d}_v\chi}(v,x,y)\,,
\\
{\rm d}_vA(v,x,y)&\equiv\frac{y^ x}{1+x}\widetilde{{\rm d}_vA}(v,x,y)\,,
\\
{\rm d}_vF_{vx}(v,x,y)&\equiv y\widetilde{{\rm d}_vF}_{vx}(v,x,y)\,,
\\
V(v,x,y)&\equiv(1-y^2)^2 (1+2 y^2)+y^4\tilde{V}(v,x,y)\,.
\end{aligned}
\end{equation}
Similar redefinitions also apply to the several independent boundary variables
\begin{equation}
\begin{aligned}
A_4(v,x)&\equiv x\,a_4(v,x)\,,
\\
p_2(v,x)&\equiv\tilde{p}_2(v,x)\,,
\\
g_2(v,x)&\equiv-\sqrt{6}+\rho(v,x)\,,
\\
U_4(v,x)&\equiv x\,u_4(v,x)\,,
\\
V_4(v,x)&\equiv v_4(v,x)-3\,.
\end{aligned}
\end{equation}

The above redefinitions are motivated by the fact that the resulting equations of motion for the tilde variables are free of singularities at $x=0$ (where the $S^2$ shrinks to zero size) and satisfy simple Dirichlet conditions as $x \to +\infty$. Additionally, the factors of $y$ are chosen so that all tilde variables obey simple Dirichlet boundary conditions at the conformal boundary. These conditions can be expressed in terms of $\chi_4$, $v_4$, $u_4$, $a_4$, $\rho$, and $\tilde{p}_2$. In particular, we have
\begin{equation}
\begin{aligned}
\tilde{A}_4(v,x,0)&=a_4(v,x)\,,
\\
\tilde{F}_{yx}(v,x,0)&=-2\tilde{g}_2(v,x)\,,
\\
\tilde{B}(v,x,0)&=0\,,
\\
\tilde{\Pi}_F(v,x,0)&=-2\rho(v,x)\,,
\\
\tilde{\Pi}^x(v,x,0)&=-u_4(v,x)\,,
\\
\tilde{U}^x(v,x,0)&=u_4(v,x)\,,
\\
\tilde{F}_{vy}(v,x,0)&=-2\rho(v,x)\,,
\\
\widetilde{{\rm d}_v \chi}(v,x,0)&=\frac{1}{2}\left[v_4(v,x)+4\chi_4(v,x)\right]\,,
\\
\widetilde{{\rm d}_v A}(v,x,0)&=-2 a_4(v,x)\,,
\\
\widetilde{{\rm d}_v F}_{vx}(v,x,0)&=-\tilde{g}_2(v,x)\,,
\end{aligned}
\label{eq:dirib}
\end{equation}

With the definitions above, it is a simple exercise to compute the relevant components of the Weyl tensor and thereby measure the tidal forces.
We first note that the following vector is tangent to future-directed, affinely parametrised null geodesics
\begin{equation}
U=e^{-2\beta(v,x,y)}y^2 \frac{\partial}{\partial y}\,.
\end{equation}

With $U$ in hand, we can express the tidal forces, as encoded by the Riemann tensor, as
\begin{subequations}
\label{eq:exacttidal}
\begin{equation}
B_{IJ}\equiv U^a U^ bR_{aI bJ}
\end{equation}
In analogy with the linearized theory, we will focus on the tracefree part of $B_{IJ}$
\begin{equation}
\widehat{B}_{IJ}\equiv U^a U^ bR_{aI bJ}-\frac{g_{IJ}}{3}U^aU^bg^{PQ}R_{aPbQ}\,,
\end{equation}
and write
\begin{equation}
\widehat{B}^{I}_{\phantom{I}J}=\mathrm{diag}\{-2\Psi,\Psi,\Psi\}
\end{equation}
with 
\begin{equation}
\Psi=e^{-4 \beta}y^4\left[\frac{3}{4}\left(\frac{\partial_yA}{A}\right)^2+\left(\frac{\partial_yA}{A}\right)\left(\partial_y\beta-\partial_y\chi\right)-\frac{1}{2}\frac{\partial_y^2A}{A}\right]\,,
\end{equation}
\end{subequations}%
 As noted earlier when discussing the linear results, $\widehat B_{IJ}$ may also be defined in terms of the Weyl tensor. One of our goals is to compare the quantity $\Psi$ with its linear counterpart $\Phi$, defined in Eq.~(\ref{eq:lineartidal}).

We now outline the integration procedure, i.e. how to evolve the system in $v$ starting from given initial data. The initial data consist of prescribing $\chi_4$, $\rho$, $u_4$, $v_4$, $\tilde{A}$, and $\tilde{F}_{yx}$ at a given time slice (for some fixed $v$), such that an apparent horizon exists at $y=1$. To achieve this, we take $\rho$, $u_4$, and $v_4+5 \chi_4$ as arbitrary functions of $x$, while $\tilde{A}$ and $\tilde{F}_{yx}$ are taken as arbitrary functions of $x$ and $y$ (these determine $a_4(v,x)$ and $\tilde{g}_2(v,x)$ via their asymptotic value at the conformal boundary see Eq.~(\ref{eq:dirib}) above). This makes physical sense: $\rho$ is setting the initial charge density, $v_4+5 \chi_4$ the initial energy density, and  $u_4$ the initial fluid velocity along $x$ with $\tilde{A}$ and $\tilde{F}_{yx}$ being the bulk dynamical degrees of freedom. We then solve for $\chi_4$ so that the apparent horizon condition is satisfied (see Eq.~(\ref{eq:appa})). This defines an elliptic problem, which we solve using a Newton–Raphson routine.
Once valid initial data are obtained, we can reconstruct $\tilde{B}$, $\tilde{\Pi}_F$, $\tilde{\Pi}^x$, $\tilde{U}^x$, $\tilde{F}_{vy}$, $\widetilde{{\rm d}_v\chi}$, $\widetilde{{\rm d}_v A}$, and $\widetilde{{\rm d}_v F}_{vx}$, together with $\tilde{V}$, $\partial_v \chi_4$, $\partial_v v_4$, $\partial_v u_4$, $\partial_v \rho$, $\partial_v \tilde{A}$, and $\partial_v \tilde{F}_{yx}$ on the time slice by solving a nested system of linear equations. All the equations share the schematic form
\begin{equation}
O_1[q_j;y]\,\partial_y q_i + O_0[q_j;y]\,q_i = S_i[q_j;y]\,,
\end{equation}
where $O_0[q_j;y]$, $O_1[q_j;y]$, and $S[q_j;y]$ are constructed from variables $q_j$ (and their derivatives in $x$ and $y$) already determined. Consequently, the variable $q_i$ can be obtained by a straightforward integration in $y$, subject to the appropriate boundary conditions. Note that no explicit $x$-derivative acts on $q_i$, which makes parallelisation along the $x$ direction straightforward.

It remains to clarify the sequence in which such a nested system of equations arises. Following \cite{Balasubramanian:2013yqa,Crisford:2017zpi} \emph{mutatis mutandis} the procedure is as follows:
\begin{enumerate}
\item From $E_{yy}=0$, we obtain a linear differential equation for $\tilde{B}$, solved with the Dirichlet boundary condition in Eqs.~(\ref{eq:dirib}).
\item From $J^v=0$, we obtain a linear differential equation for $\tilde{\Pi}_F$, also solved with the boundary condition in Eqs.~(\ref{eq:dirib}).
\item From $E_{xy}=0$, we determine $\tilde{\Pi}^x$, again imposing Eqs.~(\ref{eq:dirib}).
\item Using the definition of $\Pi^x$ in Eq.~(\ref{eq:momPi}), we then determine $\tilde{U}^x$, with boundary conditions from Eqs.(\ref{eq:dirib}).
\item From the definition of $\Pi^F$ in Eq.~(\ref{eq:momPF}), we determine $\tilde{F}_{vy}$, imposing Eqs.~(\ref{eq:dirib}).
\item The components $h^{IJ}E_{IJ}=0$ yield a regular equation for $\widetilde{{\rm d}_v \chi}$, with no singularities at $y=0$ or $y=1$. For the boundary condition, we may either impose the apparent horizon condition, given in Eq.~(\ref{eq:appa1}), or the Dirichlet condition at the conformal boundary, given in Eqs.(\ref{eq:dirib}). We adopt the apparent horizon condition and monitor Eq.~(\ref{eq:dirib}) as the system evolves, using it as an error check.
\item The coupled system $E_{xx}=0$ and $J_x=0$ is solved for $\widetilde{{\rm d}_v A}$ and $\widetilde{{\rm d}_v F}_{vx}$, using Dirichlet boundary conditions, given in Eqs.~(\ref{eq:dirib}).
\item Evaluating $E_{vy}+U^I E_{Iy}=0$ on the apparent horizon gives a linear elliptic equation for $\tilde{V}(v,x,1)$, which we solve by imposing smoothness at $x=0$ and finiteness of $\lim_{x \to +\infty}\tilde{V}(v,x,1)$.
\item With $\tilde{V}(v,x,1)$ on the horizon and $\widetilde{{\rm d}_v \chi}$ everywhere, the form of $\chi$ in terms of $\chi_4$, given in Eq.(\ref{eq:pg}), allows us to determine $\partial_v \chi_4$.
\item Knowing $\partial_v \chi_4$ and $\widetilde{{\rm d}_v \chi}$ then allows us to reconstruct $\tilde{V}(v,x,y)$ throughout the bulk.
\item From the Bianchi identity, given in Eq.~(\ref{eq:bianchi}), and the definition of $\widetilde{{\rm d}_v A}$, we determine $\partial_v \tilde{F}_{yx}$ and $\partial_v \tilde{A}$.
\item Finally, $\partial_v v_4$, $\partial_v u_4$ and $\partial_v \rho$ follow from conservation of the stress-energy tensor and charge (see Eqs.~(\ref{eqs:cons})).
\end{enumerate}
At this stage, we have determined $\partial_v \chi_4$, $\partial_v v_4$, $\partial_v u_4$, $\partial_v \rho$, $\partial_v \tilde{A}$, and $\partial_v \tilde{F}_{yx}$, which we evolve forward in $v$ using a fourth-order Runge–Kutta method.

For spatial discretisation along the holographic direction $y$, we employ a spectral element mesh with Legendre–Gauss–Lobatto nodes, and inter-element coupling is handled using a discontinuous Galerkin method with Lax–Friedrichs flux (similar techniques were used with great success in \cite{Choptuik:2017cyd}). To resolve the sharp gradients that develop over time, we implement adaptive mesh refinement along the radial direction. The primary numerical limitation stems from the formation of large \emph{radial} gradients, which require a local increase in the number of elements near the horizon and ultimately slow the simulation due to the Courant–Friedrichs–Lewy condition. Along the semi-infinite direction $x \in \mathbb{R}^+$, the system is discretised using interpolation on Laguerre nodes (see Appendix~\ref{app:lagrange}).

\section{Results\label{sec:results}}
We now present our results  evolving both linear and nonlinear perturbations.
We have considered various forms of initial data, but to present our results, we focus on initial data given by
\begin{subequations}
\begin{equation}
\tilde{A}(0,x,y) = A_1\,e^{-x},\, 
\qquad 
\tilde{F}_{yx}(0,x,y) = B_1\,e^{-x}\,,
\end{equation}
and
\begin{equation}
v_4(0,x) + 5 \chi_4(0,x) = \rho(0,x) = u_4(0,x) = 0\,.
\end{equation}
\label{eq:initialdata}
\end{subequations}
 In other words, the data at $v=0$ is given by \eqref{eq:nonlinear} with $A(0,x,y) = 1+ A_1 y^4 x e^{-x},\  U^x = 0, \ \chi = 0$.
The last three conditions in \eqref{eq:initialdata} ensure that, at $v=0$, the system starts with the energy density, charge density, and fluid velocity of an extremal RN black hole. Since $x=R^2$ in the flat boundary metric, this is like a Gaussian profile for the deformation and the rapid decay for large $x$ ensures that the system remains at zero temperature far from the deformation.  
\subsection{Linear}
Since our focus is on perturbations of extreme RN black holes, we set $\tilde{Q}= L Q/r_+^3 = \sqrt{2}$. To construct the initial data, we decompose \eqref{eq:initialdata} into harmonics $\mathbb{Y}_k$ using standard Fourier transform methods. This procedure translates the behavior of each $k$-mode (or equivalently, each $\tilde{k}\equiv Lk/r_+$ mode) into initial data for the linear problem expressed through the gauge invariant variables $Q_1(0,y)$ and $Q_2(0,y)$ \eqref{eq:Qdef}. After some algebra, we find
\begin{equation}
\begin{aligned}
&Q_1(0,y)=\frac{y e^{-\frac{\tilde{k}^2}{4}} \tilde{k}^2}{8 \sqrt{\pi }} \frac{15 \sqrt{6} A_1 y^3 \tilde{k}^2+4 B_1 (\tilde{k}^2+18 y^2)}{ \tilde{k}^2+18 y^2 (1-y^2)}
\\
&Q_2(0,y)=\frac{y^2 e^{-\frac{\tilde{k}^2}{4}} \tilde{k}^4}{12 \sqrt{\pi } }\frac{3 A_1 (\tilde{k}^2-3 y^4+3 y^2)+2 \sqrt{6} B_1 y}{\tilde{k}^2+18 y^2 (1-y^2)}
\end{aligned}
\end{equation}

If our earlier arguments about the behavior of linearized  perturbations is correct \eqref{eq:EMmodes}, we expect to see the growth of the tidal forces on the horizon (described by $\Phi(y=1)$ in \eqref{eq:lineartidal}), for each $\tilde{k}$-mode, at late times of the form
\begin{subequations}
\begin{equation}
\tilde{\Phi}_{\tilde{k}}\sim\hat{v}^{2-\alpha(\tilde{k})}
\end{equation}
where $\hat{v}={L^2 v /r_+}$,  $\tilde{\Phi}_{\tilde{k}}(\hat{v},y)\equiv \Phi_k(v,r)r_+^4/L^2$ and 
\begin{equation}
\alpha(\tilde{k})=\frac{1}{2} \left(\sqrt{5+\frac{\tilde{k}^2}{3}-4 \sqrt{1+\frac{2 \tilde{k}^2}{9}}}+1\right)\,.
\end{equation}
\end{subequations}%
This means, in particular, that the growth is maximized for $\tilde{k}=\sqrt{7/2}$, in which case
\begin{equation}
\tilde{\Phi}_{\sqrt{7/2}}\sim\hat{v}^{\frac{3}{2}-\frac{1}{2}\sqrt{\frac{5}{6}}}\approx \hat{v}^{1.04356}\,,
\end{equation}
and that for $|\tilde{k}|>2 \sqrt{14}$, the tidal forces decay at late $\hat{v}$ time.

\begin{figure}[ht]
\centering
\includegraphics[width=0.8\textwidth]{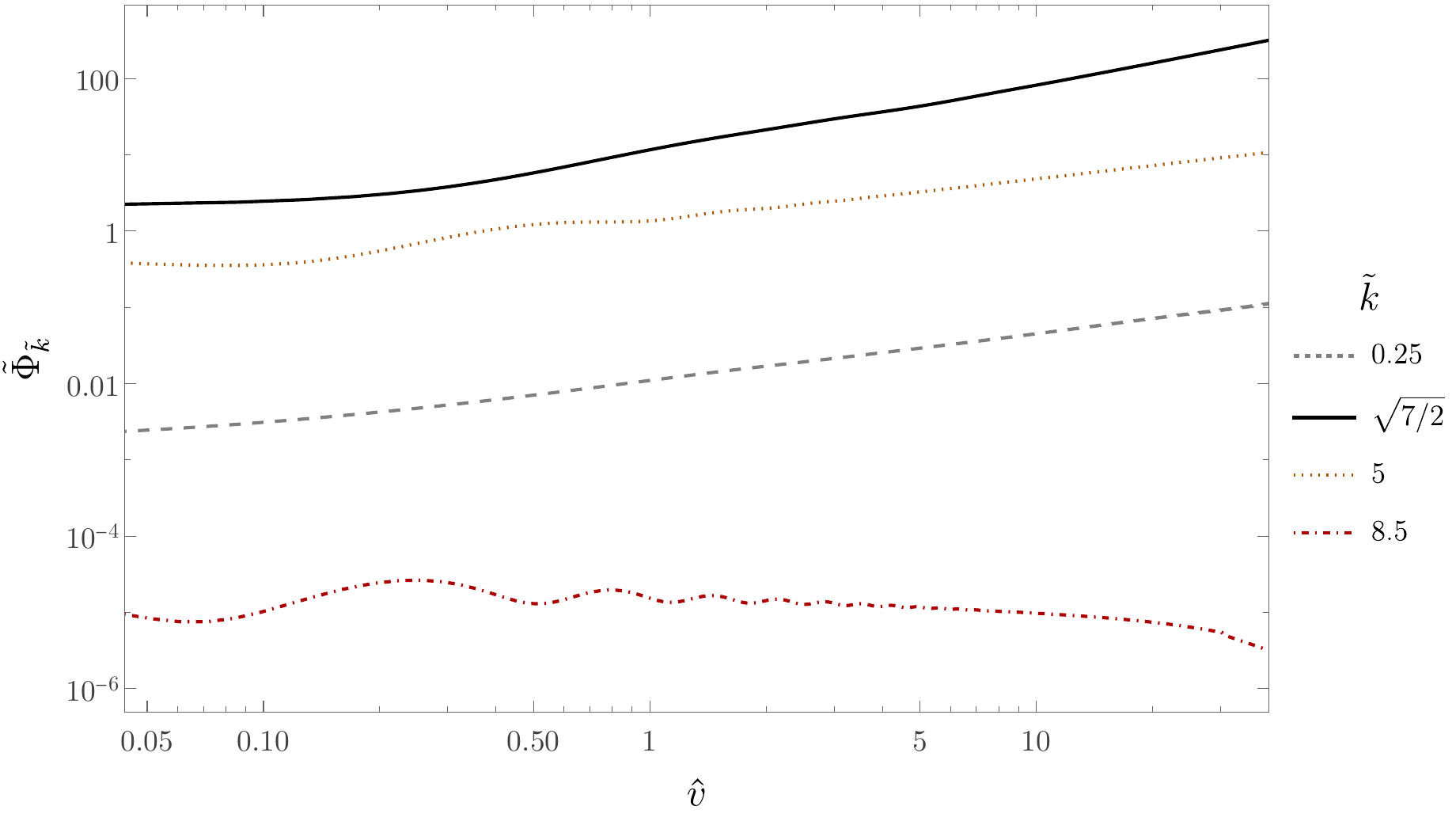}
\caption{Modes of the rescaled horizon tidal curvature $\tilde{\Phi}_{\tilde{k}}$ \eqref{eq:lineartidal} versus rescaled time $\hat{v}$ on a $\log-\log$ scale, shown for several values of $\tilde{k}$ labeled on the right, with initial data $A_1=1/2$ and $B_1=1$ in \eqref{eq:initialdata}. The late time evolution is consistent in all plots with $\hat{v}^{2-\alpha(\tilde{k})}$ at late times. At intermediate times, the system exhibits a ringdown phase, characteristic of the corresponding quasinormal mode spectrum.}
\label{fig:linear}
\end{figure}

In Fig.~\ref{fig:linear} we display $\tilde{\Phi}_{\tilde{k}}$ as a function of $\hat{v}$ on a $\log-\log$ scale, for several values of $\tilde{k}$ with $A_1=1/2$ and $B_1=1$. In all cases, the late time behaviour is very well approximated by $\tilde{\Phi}_{\tilde{k}}\sim\hat{v}^{2-\alpha(\tilde{k})}$. We have tested various choices of initial data and different values of $A_1$ and $B_1$, and the observed behavior remains unchanged. At intermediate times, the system exhibits a ringdown phase, characteristic of the corresponding quasinormal mode spectrum. This is particularly the case for large values of $\tilde{k}$, since those modes have smaller time-scales \cite{Janiszewski:2015ura}.

\subsection{Nonlinear} \label{sec:nonlinear}

 We now describe the results of our nonlinear evolution. We will show that tidal forces on the horizon grow with time, and in fact, the linear analysis reliably predicts the long-time behaviour of the non-linear evolution. This may seem surprising, since large gradients do develop over time. We will also show that the non-trivial time-evolution takes place very close to the horizon. Furthermore,  all one-point functions in the dual field theory appear to decay at late times back to their RN values.

 We have explored different choices of initial data, including polynomial decay at large $x$, and found similar results. However, the numerical simulations for initial data with polynomial decay in $x$ are more difficult to stabilise. This is likely due to the fact that our numerical discretisation along $x$, based on Laguerre functions, is well adapted to exponentially decaying functions, but is known to encounter difficulties with polynomial decay.

\begin{figure}[ht]
\centering
\includegraphics[width=0.9\textwidth]{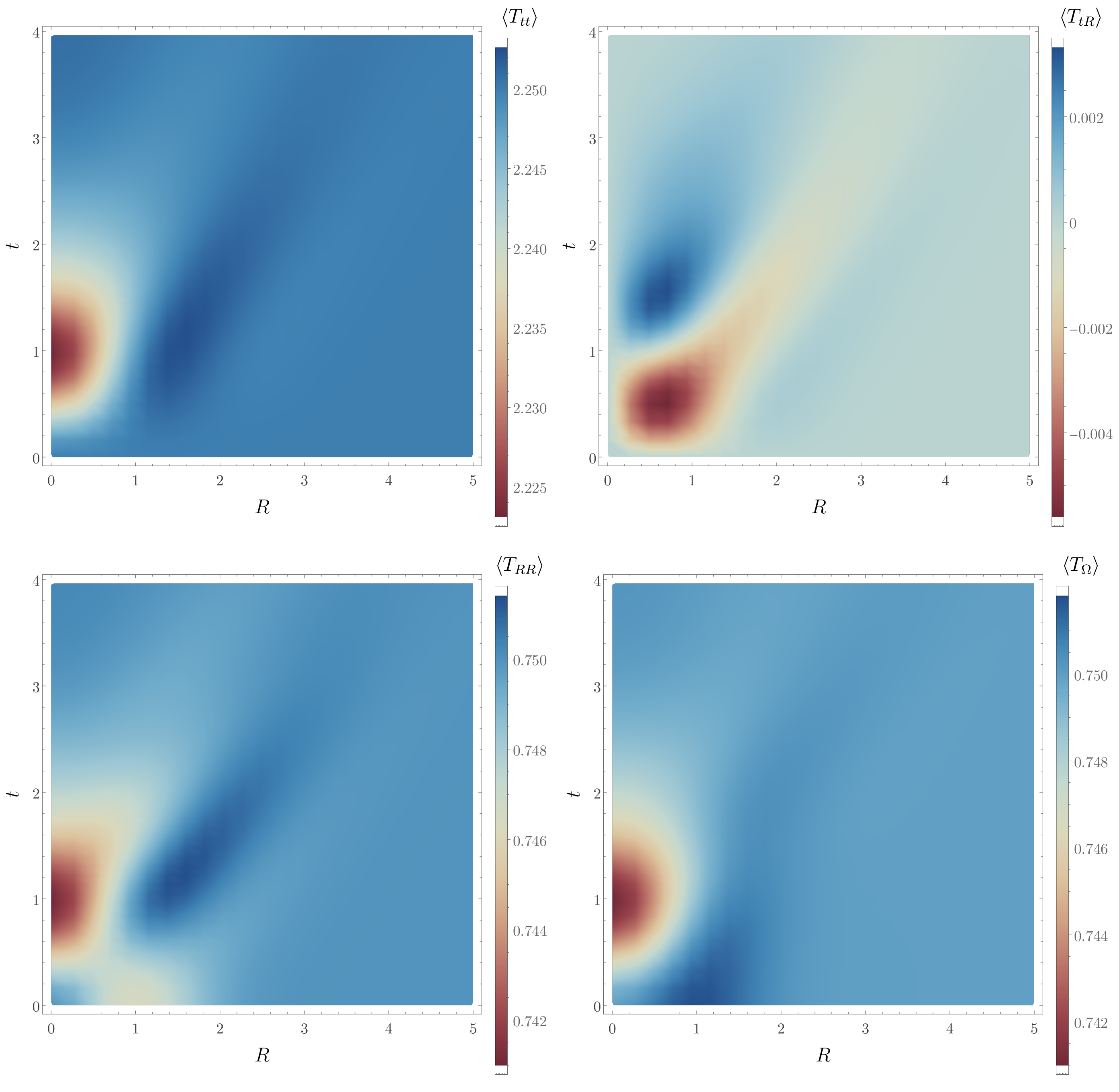}
\caption{Plots of the boundary one-point functions $\langle T_{tt}\rangle$, $\langle T_{tR}\rangle$, $\langle T_{RR}\rangle$, and $\langle T_{\Omega}\rangle$ as functions of $R$ and $t$. In all cases, the quantities begin at their extreme planar black hole values, 
 develop structure near the origin corresponding to our localized bulk deformation, and decay at late times back to the extreme planar black hole values. These plots were generated using the initial data specified in Eq.~(\ref{eq:initialdata}), with $B_1 = 2A_1 = 10^{-2}$.}
\label{fig:stress}
\end{figure}

While we tested several values of $A_1$ and $B_1$ in Eq.~(\ref{eq:initialdata}), here we fix $B_1=2A_1 = 10^{-2}$. We start by displaying the boundary one-point functions $\langle T_{tt}\rangle$, $\langle T_{tR}\rangle$, $\langle T_{RR}\rangle$, and $\langle T_{\Omega}\rangle$ as functions of $R$ and $t$ in Fig.~\ref{fig:stress}. In all cases, the quantities start from their extreme planar RN values, develop structure near the origin corresponding to our localized bulk deformation, and then 
decay at late times back to the extreme planar RN values.

\begin{figure}[ht]
\centering
\includegraphics[width=0.8\textwidth]{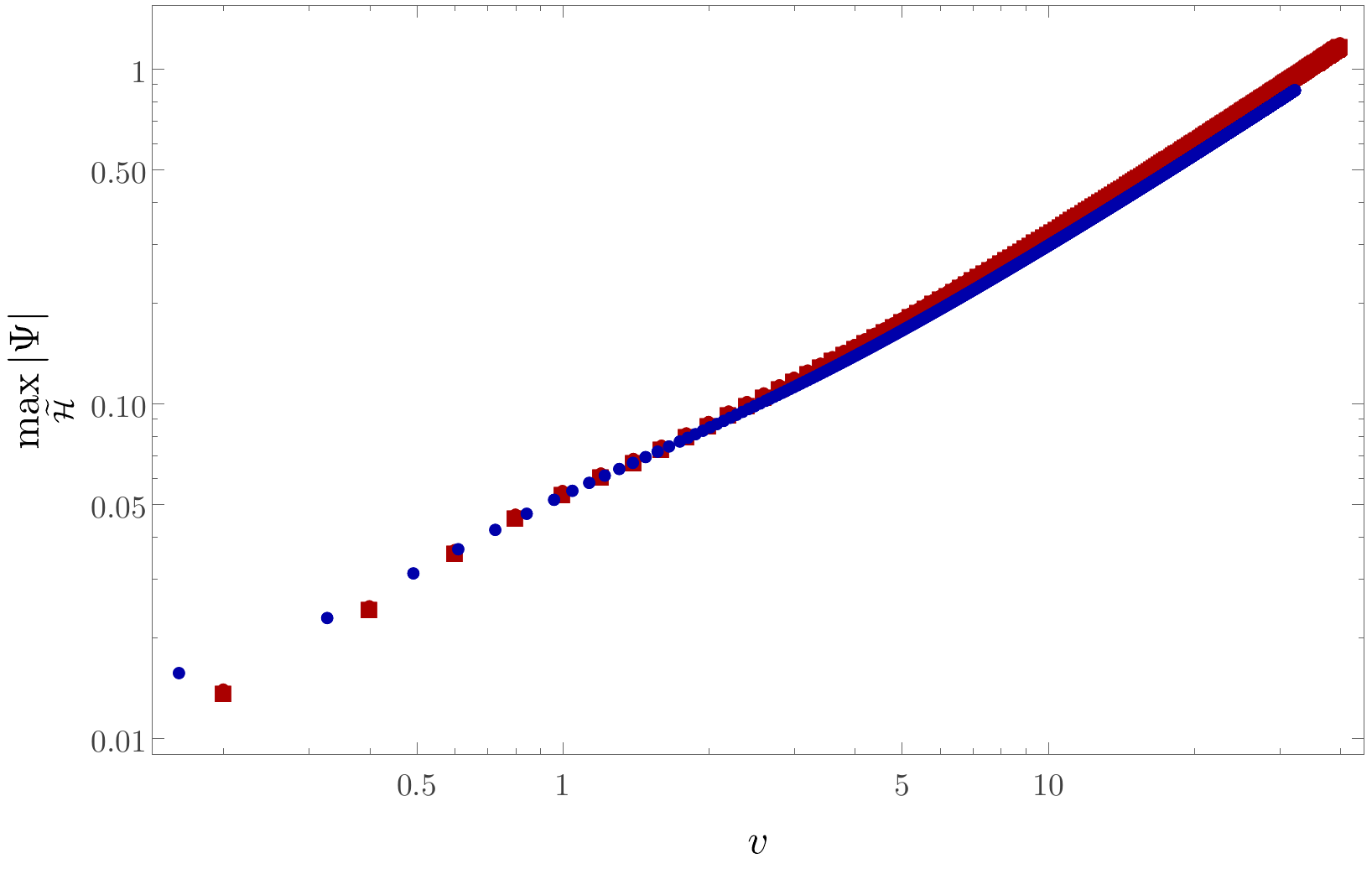}
\caption{Maximum of tidal force $|\Psi|$ \eqref{eq:exacttidal} on the apparent horizon $\widetilde{\mathcal{H}}$ (corresponding to $y=1$ in our coordinates) as a function of time $v$. Blue disks represent the nonlinear evolution with initial data $B_1=2 A_1=10^{-2}$ in \eqref{eq:initialdata}, while red squares show the linear result obtained by superimposing forty linear simulations with uniformly spaced $\tilde{k}\in[0.25,10]$. At late times the two curves have equal slope in this $\log-\log$ plot, and for sufficiently small $A_1$ and $B_1$ they coincide.}
\label{fig:growth_nl}
\end{figure}

The tidal force on an ingoing observer is given in \eqref{eq:exacttidal}.  We find that the pure trace piece decays to a small constant (consistent with being zero) at late times, so we focus on the trace-free part $\Psi$. Fig.~\ref{fig:growth_nl} shows $\max_{\widetilde{\mathcal{H}}}|\Psi|$ as a function of $v$, where $\widetilde{\mathcal{H}}$  represents the apparent horizon (recall that in our coordinates this corresponds to the surface $y=1$). While the apparent horizon is not strictly the black hole horizon, ray-tracing confirms that, for the amplitudes studied here, it coincides with the true horizon after a very short time. This plot contains two types of data. The blue disks depict the nonlinear evolution with $B_1=2A_1=10^{-2}$, whereas the red squares show the linear result obtained by Fourier transforming the initial data and superimposing forty simulations with evenly spaced $\tilde{k}$. At late times the curves behave almost identically, and for sufficiently small $A_1$ and $B_1$ they coincide. For this initial data, $\max_{\widetilde{\mathcal{H}}}|\Psi|$ grows from $\sim 0.01$ to $\mathcal{O}(1)$ by $v \sim 30$, marking the onset of the fully nonlinear regime. The growth at late times is completely consistent with the linear result $v^{1.04356}$.\footnote{ We cannot evolve long enough to see the log corrections discussed in Sec. 2.}

\begin{figure}[ht]
\centering
\includegraphics[width=0.8\textwidth]{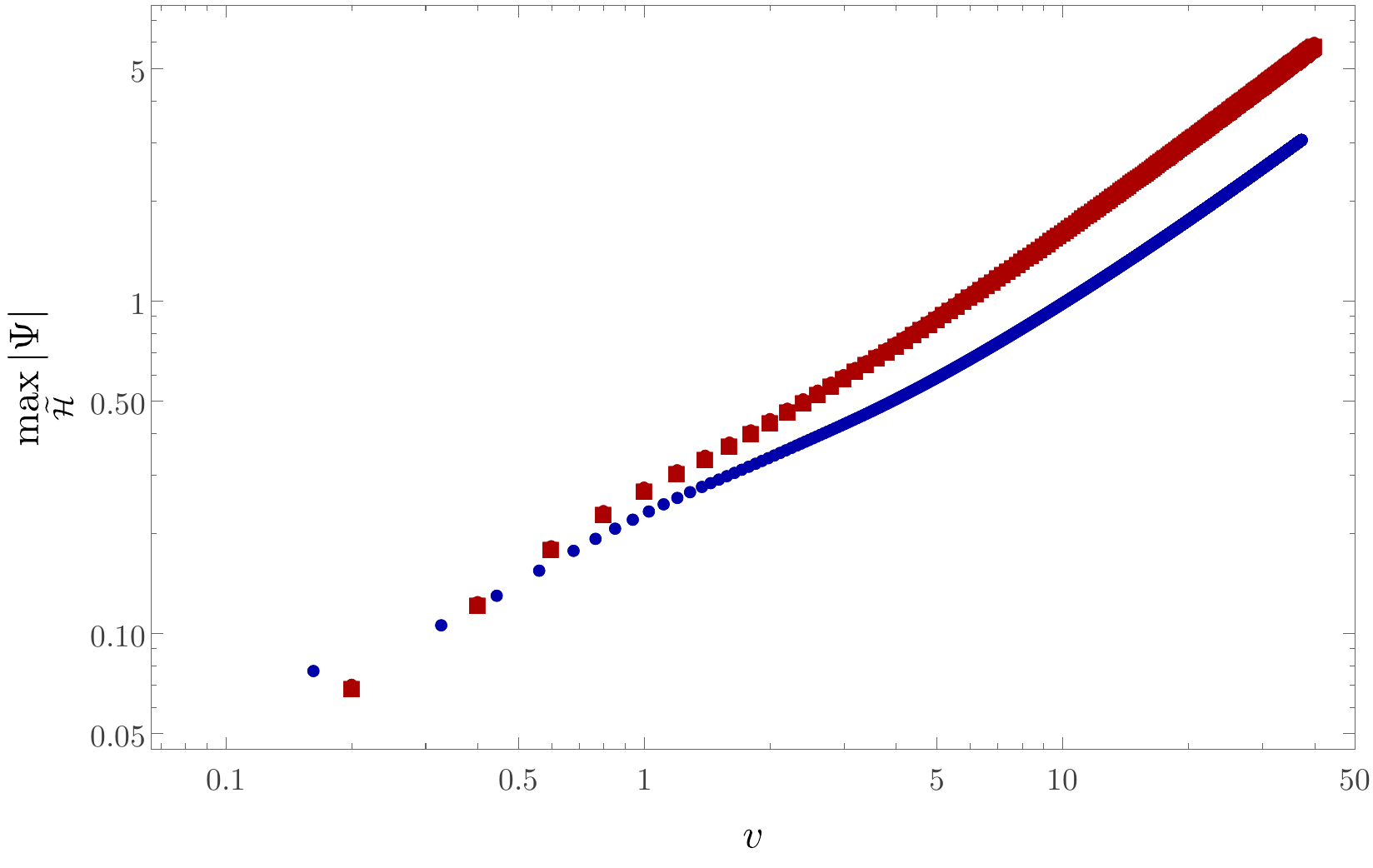}
\caption{Maximum of tidal force $|\Psi|$ \eqref{eq:exacttidal} on the apparent horizon $\widetilde{\mathcal{H}}$ as a function of $v$. Blue disks represent the nonlinear evolution with larger initial data $B_1=2 A_1=5\times 10^{-2}$ in \eqref{eq:initialdata}, while red squares show the linear result obtained by superimposing forty linear simulations with uniformly spaced $\tilde{k}\in[0.25,10]$. At late times the two curves have equal slope in this $\log$-$\log$ plot showing they grow at the same rate.}
\label{fig:growth_nl2}
\end{figure}

In Fig.~\ref{fig:growth_nl2}, the blue disks represent the nonlinear evolution with  larger initial data
$B_1 = 2A_1 = 5 \times 10^{-2}$, while the red squares show the linear result obtained by superimposing forty simulations with evenly spaced values of $\tilde{k}$. At late time the slopes of the two curves are identical and again consistent with $v^{1.04356}$. The fact that the nonlinear growth has a smaller coefficient can be understood from the intuitive picture given in the introduction: the initial localized nonlinear perturbation spreads out along the horizon and decreases its amplitude. Eventually it resembles a linear evolution, but with a smaller amplitude for the fastest growing mode than if we had used a linear analysis from the beginning. This picture also explains why the power law growth takes longer to get established when we increase the initial data.


\begin{figure}[ht]
\centering
\includegraphics[width=0.8\textwidth]{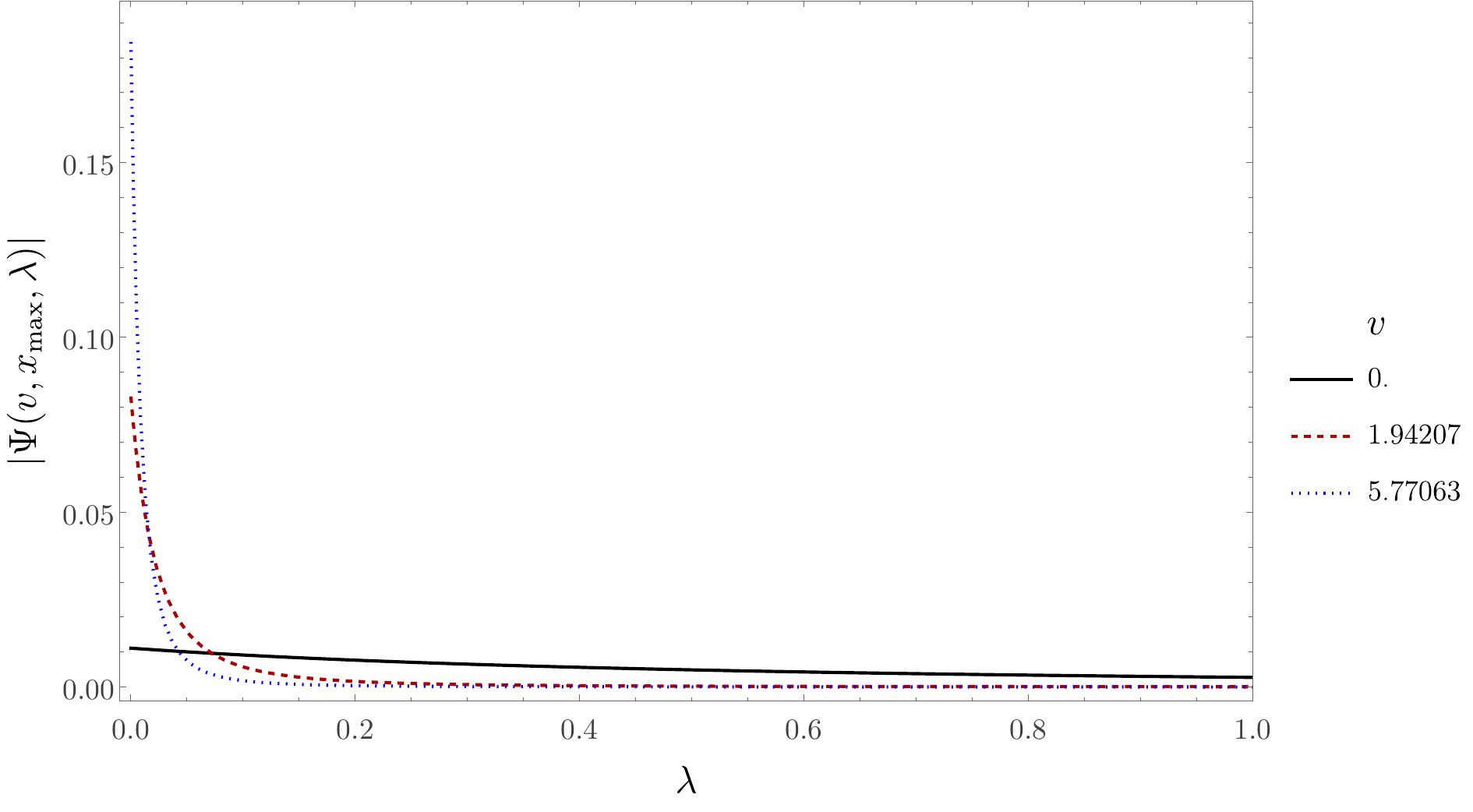}
\caption{Profiles of the tidal force $|\Psi(v, x_{\max}, \lambda)|$ along ingoing null geodesics for several values of $v$ ($\lambda$ is the affine parameter \eqref{eq:affine}), showing that the profile becomes increasingly steep near the horizon, $\lambda=0$, as $v$ increases. This plot was generated using initial data  with $B_1 = 2A_1 = 10^{-2}$ in (\ref{eq:initialdata}).}
\label{fig:detail1}
\end{figure}

We have confirmed that the growing curvature is concentrated near the horizon. 
To quantify this effect, we employed two diagnostics. For each time $v$, we first identified the point on the horizon where $|\Psi|$ reaches its maximum, 
and denoted the corresponding $x$-coordinate by $x_{\max}$. 
 We then considered the profile of $|\Psi|$ 
  along a future-directed ingoing null geodesic which reaches the horizon at $(v,x_{\max})$. The tangent vector of this geodesic is given by
\begin{equation}
U = e^{-2\beta(v,x,y)}\, y^2 \frac{\partial}{\partial y},
\end{equation}
and its affine parameter $\lambda$ is defined as
\begin{equation}\label{eq:affine}
\lambda(v,x,y) = \int_{y}^{1} \frac{e^{2\beta(v,x,\tilde{y})}}{\tilde{y}^2}\, \mathrm{d}\tilde{y}.
\end{equation}
By construction, this affine parameter is normalised so that $\lambda = 0$, and $\partial \lambda/\partial y=1/y$ near $y=0$. The resulting profiles are shown in Fig.~\ref{fig:detail1}, 
with several representative values of $v$ indicated. It is evident that, as $v$ increases, the profiles become increasingly  steep around $\lambda=0$.

\begin{figure}[ht]
\centering
\includegraphics[width=\textwidth]{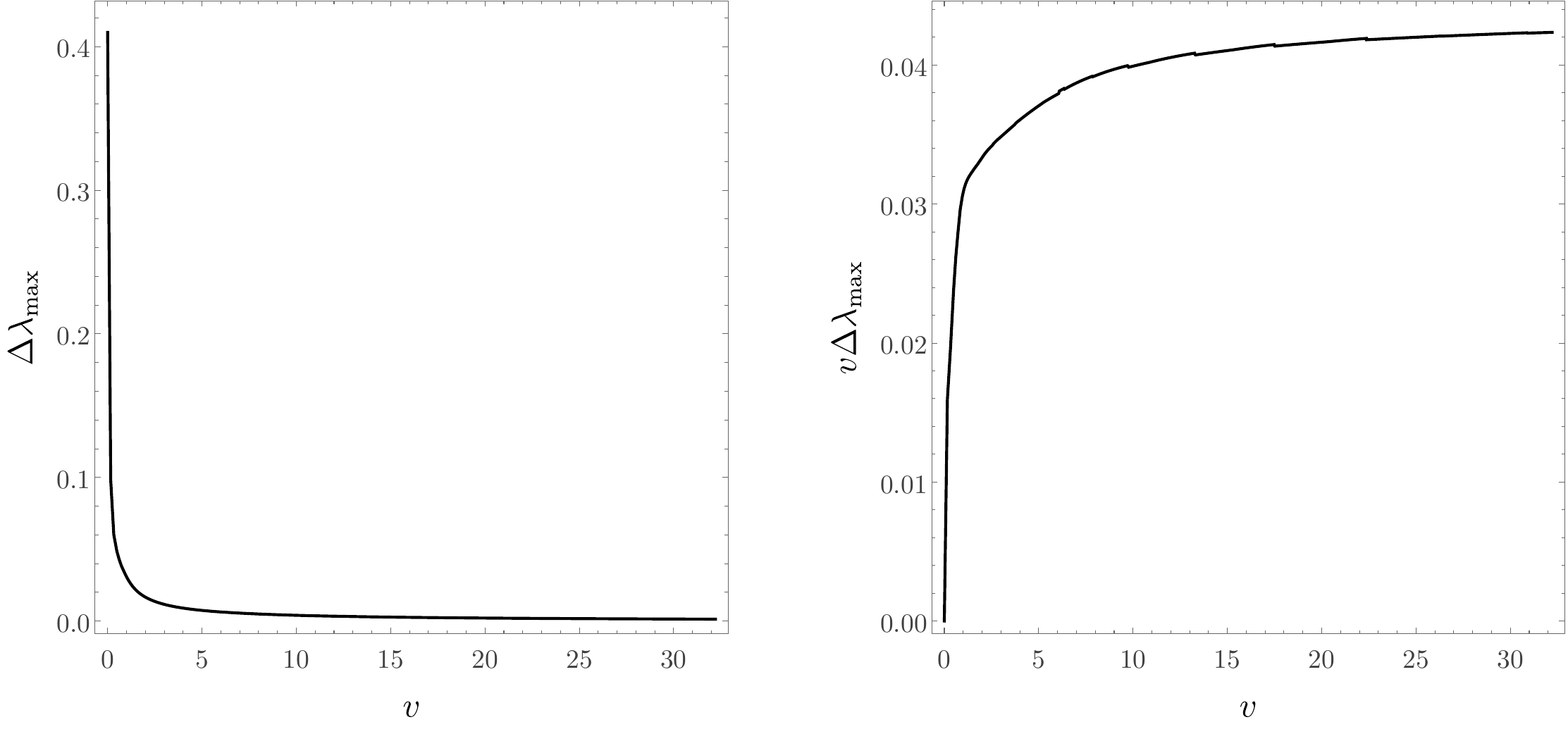}
\caption{ \textbf{Left:} $\Delta \lambda_{\max}$ as a function of $v$, where $\Delta \lambda_{\max}$ is defined as the affine-parameter distance from the horizon at which $|\Psi(v, x_{\max}, \lambda)|$ falls to half of its maximum value. As $\Delta \lambda_{\max}$ tends to zero with increasing $v$, the curvature becomes progressively localized near the horizon. \textbf{Right:} $v \Delta \lambda_{\max}$ as a function of $v$. Its rapid saturation to a constant at late times indicates that $\Delta \lambda_{\max} \sim v^{-1}$ for large $v$. Both panels are based on  initial data with $B_1 = 2A_1 = 10^{-2}$ in \eqref{eq:initialdata}.}
\label{fig:detail2}
\end{figure}

As a second diagnostic, we define $\Delta \lambda_{\max}$ as the value of $\lambda$ at which $|\Psi(v, x_{\max}, \lambda)|$ decreases to half of its maximum. The left panel of Fig.~\ref{fig:detail2} shows $\Delta \lambda_{\max}$, which tends to zero as $v$ increases, indicating that the curvature becomes increasingly localized near the horizon. The right panel shows $v\,\Delta \lambda_{\max}$, which approaches a constant at large $v$, demonstrating that $\Delta \lambda_{\max} \sim v^{-1}$ at late times. 
This behavior can be read off from the linear ansatz \eqref{eq:ansatz_late_time}. There exists a constant $y_0 > 0$ such that $h(y_0) = \frac{1}{2} h(0)$. Since the argument of $h$ is $\rho v$ and $\rho = 0$ corresponds to the horizon, it follows that the tidal forces decrease to half of its value at $\rho = \frac{y_0}{v}$. Of course, we expect this analysis to be valid only at sufficiently late times, hence we see this behavior only asymptotically.

One might wonder whether $\max_{\widetilde{\mathcal{H}}}|\Psi|$ at large $v$ 
is dominated by the special value $\tilde{k} = \sqrt{7/2}$, where 
$-\alpha(\tilde{k})$ attains its maximum. This is the fastest growing mode of the linearized theory.
Since the intrinsic geometry of the apparent horizon rapidly approaches that of an 
extreme RN black hole, we can exploit its translational invariance along 
the $x$--direction and decompose $\Psi(v, x, y)$ into partial 
$\tilde{k}$--modes on the apparent horizon. Referring to the linearized counterpart in 
Eq.~(\ref{eq:linear}), we see that the natural decomposition is given by
\begin{equation}
\Psi^{\widetilde{\mathcal{H}}}(v,\tilde{R}) \equiv \Psi(v,\sqrt{\tilde{R}},1) 
= \int_0^{+\infty} \left[\Grad_{\tilde{R}}\Grad_{\tilde{R}} \mathbb{Y}_{\tilde{k}} 
+ \frac{\tilde{k}^2}{3}\mathbb{Y}_{\tilde{k}} \right] 
\Psi^{\widetilde{\mathcal{H}}}_{\tilde{k}}(v)\, \mathrm{d}\tilde{k}\,,
\end{equation}
from which we aim to extract the coefficients 
$\Psi^{\widetilde{\mathcal{H}}}_{\tilde{k}}(v)$. 
Numerically, we have direct access to 
$\Psi^{\widetilde{\mathcal{H}}}(v,\tilde{R})$, 
and our task is to recover $\Psi^{\widetilde{\mathcal{H}}}_{\tilde{k}}(v)$ 
by inverting this integral relation.

This might look like a hard numerical task, but it turns out to be feasible. First, define
\begin{equation}
\mathcal{D}_{\tilde{R}}^2\Psi^{\widetilde{\mathcal{H}}}(v,\tilde{R})\equiv \partial^2_{\tilde{R}}\Psi^{\widetilde{\mathcal{H}}}(v,\tilde{R})+\frac{5}{\tilde{R}}\partial_{\tilde{R}}\Psi^{\widetilde{\mathcal{H}}}(v,\tilde{R})+\frac{3}{\tilde{R}^2}\Psi^{\widetilde{\mathcal{H}}}(v,\tilde{R})\,.
\end{equation}
Using the properties of the mode functions $\mathbb{Y}_{\tilde{k}}(\tilde{R})$, we obtain
\begin{equation}
\mathcal{D}_{\tilde{R}}^2\Psi^{\widetilde{\mathcal{H}}}(v,\tilde{R})
= \int_0^{+\infty} \tilde{k}^2 \, \mathbb{Y}_{\tilde{k}}(\tilde{R})
\left[\frac{2}{3}\tilde{k}^2 \Psi^{\widetilde{\mathcal{H}}}_{\tilde{k}}(v)\right] 
\, \mathrm{d}\tilde{k}\,,
\label{eq:trans}
\end{equation}
where the left-hand side can be directly obtained from our numerical simulations.

At this point, we recall that the mode functions 
$\mathbb{Y}_{\tilde{k}}(\tilde{R})$ can be expressed in terms of a 
half-order Bessel function as
\begin{equation}
\mathbb{Y}_{\tilde{k}}(\tilde{R})
= \sqrt{\frac{\pi}{2}} \, \frac{J_{1/2}(\tilde{k}\tilde{R})}{\sqrt{\tilde{k}\tilde{R}}}\,.
\end{equation}
With this representation, Eq.~(\ref{eq:trans}) takes the form of a 
Hankel transform, which can be inverted using the formulas provided 
in Appendix~\ref{app:appbessel}.

In Fig.~\ref{fig:transform}, we plot $\Psi^{\widetilde{\mathcal{H}}}_{\tilde{k}}(v)$ 
as a function of $v$ and $\tilde{k}$ for initial perturbation parameters  $B_1 = 2 A_1 = 5\times10^{-2}$. To guide the eye, we indicate 
$\tilde{k} = \sqrt{7/2}$ with a black dashed line. 
The figure clearly shows that $\Psi^{\widetilde{\mathcal{H}}}_{\tilde{k}}(v)$ 
grows predominantly at $\tilde{k} = \sqrt{7/2}$, confirming the significance 
of this mode.

\begin{figure}[ht]
\centering
\includegraphics[width=0.6\textwidth]{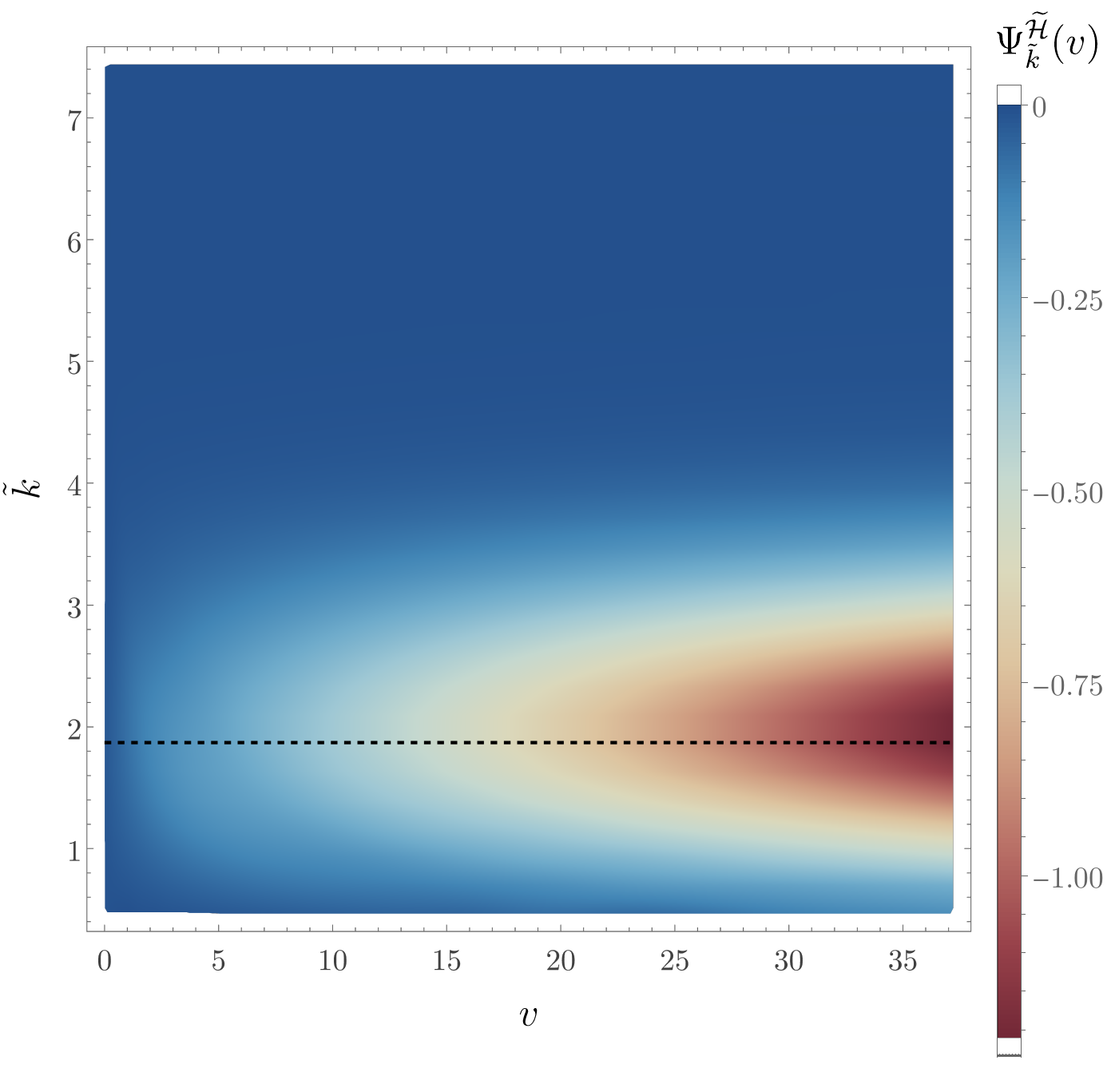}
\caption{Mode decomposition of $\Psi(v)$  along the horizon. The quantity $\Psi^{\widetilde{\mathcal{H}}}_{\tilde{k}}(v)$ is plotted as a function of  $v$ and $\tilde{k}$. The black dashed line indicates $\tilde{k} = \sqrt{7/2}$, 
highlighting that the mode amplitude grows predominantly at this special value. The initial perturbation parameters are $B_1 = 2 A_1 = 5\times10^{-2}$ in \eqref{eq:initialdata}.}
\label{fig:transform}
\end{figure} 

\begin{figure}[ht]
\centering
\includegraphics[width=0.6\textwidth]{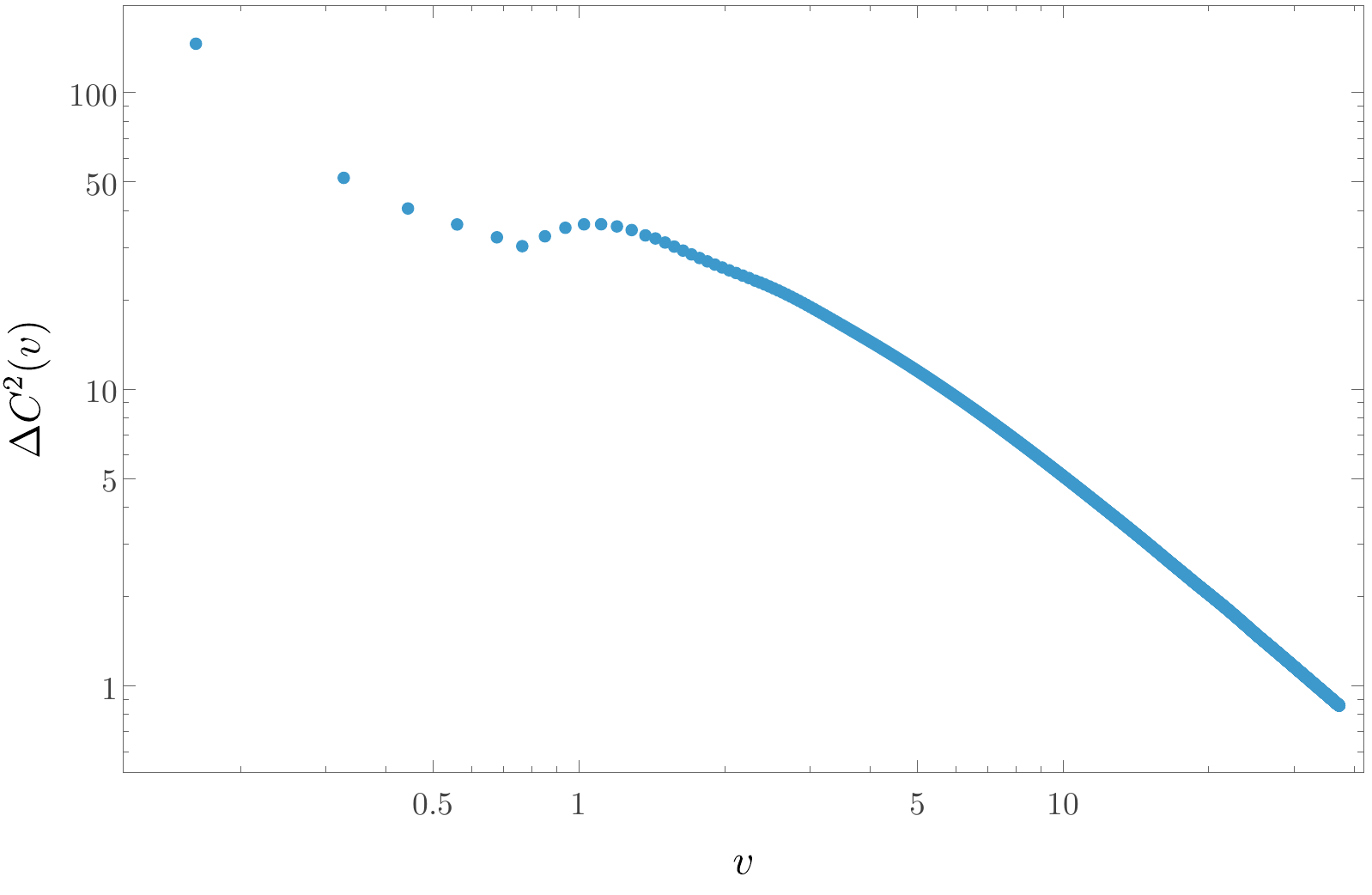}
\caption{The deviation of the Weyl curvature squared decays on the event horizon. 
We plot the maximum over the horizon of the absolute difference 
$\Delta C^2(v) \equiv \max_{\widetilde{\mathcal{H}}} |C^2 - C^2_{\rm RN}|$ as a function of advanced time $v$ on a $\log$-$\log$ scale. The initial perturbation parameters are $B_1 = 2 A_1 = 5\times10^{-2}$ in \eqref{eq:initialdata}.  
}

\label{fig:weyl}
\end{figure}

One may ask whether curvature invariants grow with time.  Our numerical results indicate they do not.  Consider, for example,
\begin{equation}
C^2 \equiv L^4\,C_{abcd}C^{abcd},
\end{equation}
where $C_{abcd}$ are the components of the Weyl tensor and $L$ is the AdS length scale used to make $C^2$ dimensionless.  We examine the behaviour of $C^2$ along the apparent horizon $\widetilde{\mathcal{H}}$ relative to its extreme RN value.  For an extremal RN black hole one has $C^2_{\mathrm{RN}} = 288$.  In Fig.~\ref{fig:weyl} we show, on a $\log$-$\log$ plot, the maximum over the horizon of the absolute difference
\begin{equation}
\Delta C^2(v) \equiv \max_{\widetilde{\mathcal{H}}} \big|\,C^2 - C^2_{\mathrm{RN}}\,\big|
\end{equation}
as a function of advanced time $v$ for initial perturbation parameters $B_1 = 2A_1 = 5\times10^{-2}$.  At late times $\Delta C^2(v)$ decays with $v$.  

\section{Discussion}
We have shown that $SO(3)$ invariant, finite energy deformations of the extremal planar black hole in AdS$_5$ lead to curvature growing without bound on the horizon. Furthermore, the late time behavior agrees with a linearized analysis. We expect similar behavior for generic finite energy deformations, since the intuitive picture is that the initial localized deformation spreads out over the horizon with decreasing amplitude. Eventually the linearized approximation is applicable and the mode with the fastest growing curvature will dominate. Curvature scalars do not grow, but the physical tidal forces felt by infalling observers increase to arbitrarily large values.  One can view this as  a frozen firewall, since it is a zero temperature analog of the firewall that might arise when a black hole evaporates. We expect similar behavior in all dimensions $D \ge 4$ and for hyperbolic black holes as well as planar. 

For convenience, we computed the tidal forces for ingoing null geodesics, but it is clear that we will get similar results for ingoing timelike geodesics (since their tangent vector always includes a component along the ingoing null direction). For null geodesics there is no preferred normalization, but we can normalize those geodesics to have unit energy at any radius. We can even start with the perturbed near horizon geometry and find growing tidal forces. 

If we were to add finite energy deformations to a near-extremal (rather than extremal) planar black hole, we expect the curvature to grow for a while  (depending on the deviation from extremality) but eventually to decay back to the original planar black hole values.

 Finite energy deformations of a compact extremal horizon can remove the horizon completely \cite{Murata:2013daa,Angelopoulos:2024yev}, but that is not possible for noncompact horizons. For example, with Minkowski space on the boundary, even in pure AdS there is a Poincar\'e horizon.  

Since the Poincar\'e horizon is also a noncompact extremal horizon, it is interesting to ask if  there is an analogous instability there. This is important for determining the stability of extremal nondilatonic black branes. For linear perturbations, it has been shown that the answer is yes, but it is a weaker instability \cite{Chen:2025sim}. The curvature itself does not diverge, but a sufficient number of transverse derivatives of the curvature will grow. The asymptotically flat region is important for interpreting this result. If one just considers the near horizon geometry, AdS$_{m} \times  S^n$, there is no preferred Poincar\'e patch. Linear perturbations remain smooth for all global time, and the growth of transverse derivatives on a Poincar\'e horizon is a purely coordinate effect.

One might ask whether one-loop quantum corrections significantly change our story.  For spherical black holes it has been shown that there is a Schwarzian mode which is not suppressed at low temperature.\footnote{The Schwarzian mode is usually analyzed in Euclidean signature, and its role for Lorentzian spacetimes is unclear. However the following argument shows that there is no analogous mode for planar black holes.} One important consequence is that the (nonsupersymmetric) black hole entropy is no longer given by the horizon area, and goes to zero in the extremal limit.  For planar black holes there is no analogous Schwarzian mode. The easiest way to see this is to work in the near-horizon geometry at $T=0$. One can write down exact zero-modes (which get regulated by temperature and give rise to Schwarzian corrections). However, if the black hole is non-compact, these modes are not normalizable, and do not contribute. So this is a qualitative difference between a very large torus and a plane. In the former case, there is a Schwarzian correction but you need to go to very small temperatures to see it. There is also a qualitative difference between a large torus and the plane in how general our result is.  For a large toroidal horizon, generic finite energy excitations will cause the curvature to grow for a while, but eventually the solution will  settle down to a nonextremal black hole. Fine tuning would be required to approach an extremal horizon and keep the curvature growing indefinitely.

Going from one-loop corrections to full holographic quantum gravity, one might wonder what is the dual CFT interpretation of the (nonlinear) instability of planar extremal black holes in AdS$_5$ that we have discussed. To answer this, recall that the standard dual theory, $\mathcal{N}=4$ Super-Yang-Mills, describes string theory in asymptotically AdS$_5 \times S^5$ spacetimes. In the supergravity limit, this bulk theory can be dimensionally reduced to $D=5$ Einstein-Maxwell, but it also includes neutral and charged scalars which cause the planar charged black hole to become unstable at low temperature \cite{Gladden:2024ssb}. The neutral scalars condense at a higher temperature than the usual superconducting instability caused by the charged scalars \cite{Buchel:2025ves}. Once the scalar fields turn on, the extremal limit is probably singular even without finite energy excitations \cite{Horowitz:2009ij}. So our results are not generic in standard holography since they require fine-tuning these  scalars to stay  zero, but they are generic in pure Einstein-Maxwell-AdS theory. It is possible that other holographic models will include $D=5$ Einstein-Maxwell without scalars that become unstable. One could then explore the dual CFT interpretation of the growing curvature.

 Since we have shown that perturbed extremal black holes become highly curved in finite time, it is natural to ask whether it constitutes a violation of weak cosmic censorship. To answer this, we must ask about the spatial extent of the curvature away from the horizon. We argued that the curvature remains of the same order of magnitude for a proper distance of the order $\rho \sim \frac{r_+^2}{v}$ (the factors of $r_+$ are inserted for dimensional purposes.) On the other hand, the curvature on the horizon is Planckian when
\begin{subequations}
    \begin{equation}
    \left(
\frac{v}{r_+}
    \right)^{2-\alpha} \sim \left(\frac{r_+}{L_{\textrm{Pl}}}\right)^2,
\end{equation}
and so on the time-scales of order 
\begin{equation}
    v \sim r_+^{\frac{4-\alpha}{2-\alpha}} L_{\textrm{Pl}}^{\frac{-2}{2-\alpha}}
\end{equation}
and it remains Planckian up to a distance
\begin{equation}
    \rho \sim L_{\textrm{Pl}}^{\frac{2}{2-\alpha}} r_+^{\frac{-\alpha}{2-\alpha}}.
\end{equation}
Since $\alpha >0$, it follows that this distance is always subplanckian. As such, even though the singularity leads to real effects (it could destroy an infalling observer, for instance), they are not observable from very far away. As such, it does not violate any version of  weak cosmic censorship.
\end{subequations}

\section*{Acknowledgements}
We thank Harvey~Reall for his collaboration at the early stages of this project and for many insightful discussions. J.~E.~S. has been partially supported by STFC consolidated grant ST/X000664/1 and by Hughes Hall College. G.~H. and M.~K. were supported in part by NSF grant PHY-2408110, and G.~H. was also supported by Simons Foundation International  and the Simons Foundation through Simons Foundation grant SFI-MPS-BH-00012593-08. This work was performed in part at the Aspen Center for Physics, which is supported by a grant from the Simons Foundation (1161654, Troyer).
\appendix

\section{The scalar toy-model \label{app:scalar}}
In this appendix we show that the near horizon analysis of a scalar field propagating on a fixed extremal planar RN black hole in Sec. \ref{sec:toy_model} accurately reproduces the late time behavior of the evolution on the full asymptotically AdS$_5$ spacetime.
We focus on the case with $\mu^2 L^2 = -3$ and $n = 3$. The corresponding scalar field is dual to an operator in the boundary CFT 
with conformal dimension $\Delta = 3$, and such fields commonly appear in consistent truncations of supergravity (see for instance \cite{Bena:2018vtu}).

We write our background metric as
\begin{equation}
\begin{aligned}
&{\rm d}s^2=-f(r){\rm d}v^2+2{\rm d}v\,{\rm d}r+r^2\left({\rm d}R^2+R^2{\rm d}\Omega_2^2\right)\,,
\\
&A=\frac{\sqrt{3}\,Q}{r^2}{\rm d}v\,,
\end{aligned}
\end{equation}
with $R>0$ and
\begin{equation}
f(r)=\frac{r^2}{L^2}-\frac{2M}{r}+\frac{Q^2}{r^4}\,.
\end{equation}

To proceed, recall that the black hole radius $r_+$ is defined by 
\begin{equation}
f(r_+)=0 \;\;\Rightarrow\;\; M=\frac{r_+^2}{L^2}\,(1+\tilde{Q}^2)\,,
\end{equation}
where $\tilde{Q}\equiv L Q/r_+^3$, with extremality at $\tilde{Q}^2=2$. We then perform the coordinate redefinitions
\begin{equation}
r=\frac{r_+}{y}\,, \qquad v=\frac{L^2}{r_+}\,\hat{v}\quad\text{and}\quad R=\frac{L \sqrt{x}}{r_+}\,.
\label{eq:compact2}
\end{equation}
The background metric becomes
\begin{equation}
{\rm d}s^2=\frac{L^2}{y^2}\left[-(1+2y^2)(1-y^2)^2{\rm d}\hat{v}^2-2\,{\rm d}\hat{v}\,{\rm d}y+\frac{{\rm d}x^2}{4x}+x\,{\rm d}\Omega_2^2\,,
\right]\,,
\end{equation}
where we set $\tilde{Q}^2=2$.

We will also restrict attention to initial data that is $SO(3)$-symmetric, i.e., the data has non-trivial dependence only on $(v, x, y)$. In these coordinates, the equation of motion for the scalar field $\phi$ reads
\begin{subequations}
\begin{equation}
\partial_y {\rm d}_{\hat{v}}\phi-\frac{3}{2y}{\rm d}_{\hat{v}}\phi-2x \partial_{x}^2\phi-3\partial_x \phi-\frac{3}{2y^2}\phi+\frac{3}{4y}(1-y^2)^2(1+2y^2)\partial_y \phi=0\,,
\end{equation}
where we defined
\begin{equation}
{\rm d}_{\hat{v}}\phi\equiv \partial_{\hat{v}}\phi-\frac{(1-y^2)^2(1+2y^2)}{2}\partial_y \phi\,.
\end{equation}
\end{subequations}
To integrate the equations, we prescribe $\phi$ on a given time slice, solve the above equation with appropriate boundary conditions at the conformal boundary (located at $y = 0$), and obtain ${\rm d}_{\hat{v}} \phi$ everywhere in the bulk. Once this quantity is determined, we compute $\partial_{\hat{v}} \phi$ and evolve the system forward using an explicit fourth-order Runge-Kutta integration scheme. It remains to describe the boundary conditions we have chosen. For $\mu^2L^2=-3$, there are two possible behaviors for $\phi$ close to the conformal boundary
\begin{equation}
\phi\approx y\,\phi_1(\hat{v},x)\left[1+\mathcal{O}(y)\right]+y^3\,\phi_3(\hat{v},x)\left[1+\mathcal{O}(y)\right]\,.
\end{equation}
We choose standard boundary conditions for $\phi$, in which case we demand $\phi= \phi_3(\hat{v},x)y^3+\mathcal{O}(y^4)$. These imply that
\begin{equation}
{\rm d}_{\hat{v}}\phi=-\frac{3}{2}\phi_3(\hat{v},x)y^2+\mathcal{O}(y^3)\,.
\end{equation}

To solve the equations, we then introduce two new auxiliary variables
\begin{equation}
\widetilde{{\rm d}_{\hat{v}}\phi}\equiv y^{-2}{\rm d}_{\hat{v}}\phi\quad\text{and}\quad \widetilde{\phi}\equiv y^{-3}\phi\,,
\end{equation}
in terms of which the boundary conditions simply read
\begin{equation}
\widetilde{{\rm d}_{\hat{v}}\phi}(\hat{v},0,x)=-\frac{3}{2}\widetilde{\phi}(\hat{v},0,x)\,.
\end{equation}

As initial data, we take
\begin{equation}
\widetilde{\phi}(0,y,x)=e^{-x}(1-y^2)^j\quad\text{with}\quad j=0,2\,.
\label{eq:initialdatascalar}
\end{equation}
At the conformal boundary,  this corresponds to a Gaussian pulse centered around $R=0$ (recall that $x\propto R^2$).

We discretize the spatial directions using a spectral element mesh with Legendre-Gauss-Lobatto nodes for the spatial discretization along the holographic direction $y$, combined with adaptive mesh refinement along the radial direction. For the semi-infinite direction $x \in \mathbb{R}^+$, the system is discretized using interpolation on Laguerre nodes.

For all the classes of initial data that we have tried, we find that $\widetilde{\phi}$ decays as $\hat{v}^{-1/2}\log^{-3} \hat{v}$ at late times, and that $\partial_y \widetilde{\phi}$ grows as $\hat{v}^{1/2}\log^{-3} \hat{v}$, as expected from the near horizon analysis.

For the particular values of $m^2L^2=-3$ and $n=3$, we can do some of the integrals appearing in the near horizon analysis at the end of section \ref{sec:toy_model} without resorting to the saddle-point approximation since\footnote{Note that we are defining our $k$ with respect to A.4, which already factors out $r_+/L$. This is what we call $\tilde{k}$ in section 2.3, just below \eqref{eq:Qdef}.}
\begin{equation}
\alpha(k)=\frac{1}{2}+\frac{k}{2\sqrt{3}}\,.
\end{equation}
Consider the case $ j = 0$ in Eq.~(\ref{eq:initialdatascalar}). If we choose
\begin{equation}
\tilde{\phi}_{\vec{k}}=\frac{1}{8\pi\sqrt{\pi}}e^{-\frac{k^2}{4}}\,,
\label{eq:gaussian2}
\end{equation}
then for $p=0$ in Eq.~(\ref{eq:near}), and under our symmetry assumptions, we find 
\begin{equation}
\phi(\tilde{v},\rho=0,x) = \frac{1}{2\sqrt{\pi}}\int_0^{+\infty} k^2\,e^{-\frac{k^2}{4}}\,e^{-\alpha(k)\,\log \tilde{v}}\frac{\sin k \sqrt{x}}{k\sqrt{x}}\mathrm{d}k\,.
\end{equation}
where $\tilde v$ is the time in the near horizon geometry.
Note that for $\tilde{v}=1$, we find
\begin{equation}
\phi(1,\rho=0,x)=e^{-x}\,,
\end{equation}
just like our initial data with $j=0$. Numerically, we observe that the maximum value of $\phi$ is always located at the origin, i.e. $x=0$. As such, we can evaluate
\begin{equation}
\begin{aligned}
\phi(\tilde{v},\rho=0,0)\equiv \phi_{\rm NH}(\tilde{v}) & = \frac{1}{2\sqrt{\pi}}\int_0^{+\infty} k^2\,e^{-\frac{k^2}{4}}\,e^{-\alpha(k)\,\log \tilde{v}}\mathrm{d}k
\\
&=\frac{1}{\sqrt{\tilde{v}}}\left[e^{\frac{1}{12} \log ^2\tilde{v}} \text{erfc}\left(\frac{\log \tilde{v}}{2 \sqrt{3}}\right) \left(1+\frac{1}{6} \log^2\tilde{v}\right)-\frac{\log
   \tilde{v}}{\sqrt{3 \pi }}\right]\,.
\end{aligned}
\end{equation}
where $\text{erfc}(z)$ is the complementary error function. Expanding at late times, reproduces Eq.~(\ref{eq:saddlepoinex}), which was obtained using the saddle-point approximation. Over time, the behavior of our Gaussian pulse should match that shown above. However, both the value of $\tilde{v}$ at which this matching commences and the corresponding relative amplitude of the mode remain unknown. To account for this effect, we will fit our numerical data to
\begin{equation}
a_0\,\phi_{\rm NH}\left(\frac{\hat{v}}{\hat{v}_0}\right)\,,
\label{eq:fit}
\end{equation}
with $a_0$ and $\hat{v}_0$ being the fitting parameters.

\begin{figure}[ht]
\centering
\includegraphics[width=0.6\textwidth]{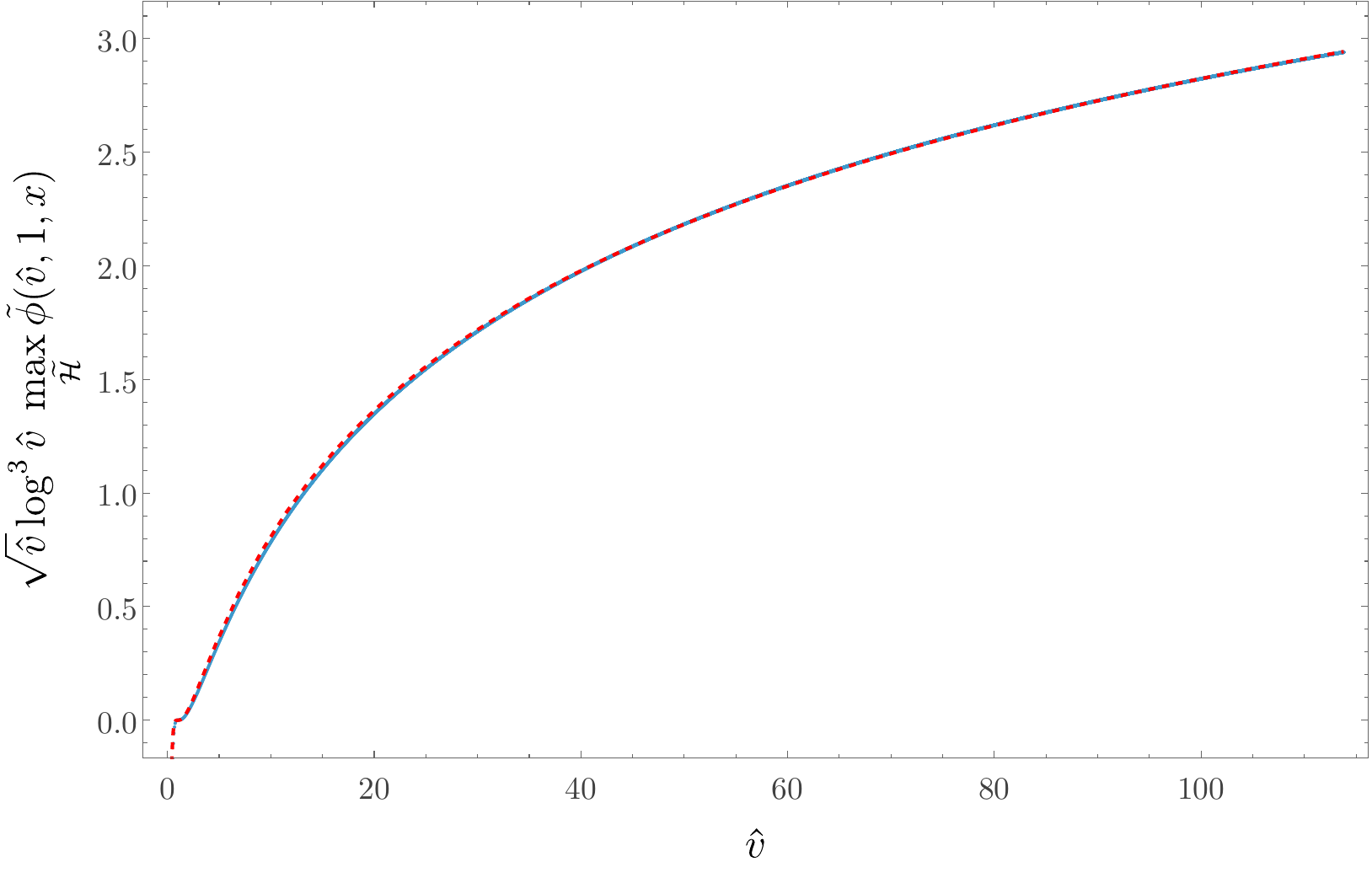}
\caption{Plot of $\sqrt{\hat{v}}\,\log^3 \hat{v}\, \max_{\widetilde{\mathcal{H}}} \tilde{\phi}(\hat{v},1,x)$ as a function of $v$ for the initial data specified in Eq.~(\ref{eq:initialdatascalar}) with $j=0$. The blue disks represent the numerical data, while the dashed red curve shows the $\chi^2$ best-fit to the analytic solution \eqref{eq:fit} 
with parameters $a_0 \approx 1.94302$ and $\hat{v}_0 \approx 0.114582$. Excellent agreement is observed for $\hat{v} \gtrsim 20$.}
\label{fig:scalar0}
\end{figure}

In Fig.~\ref{fig:scalar0}, we plot $\sqrt{\hat{v}}\,\log^3 \hat{v}\, \max_{\widetilde{\mathcal{H}}} \tilde{\phi}(\hat{v},1,x)$ as a function of $\hat{v}$ for the initial data given in Eq.~(\ref{eq:initialdatascalar}) with $j=0$. The numerical data are shown as blue disks, while the dashed red curve represents a $\chi^2$ best-fit with $a_0 \approx 1.94302$ and $\hat{v}_0 \approx 0.114582$. The agreement is excellent for $\hat{v} \gtrsim 20$.

We have also examined the behavior of $\max_{\widetilde{\mathcal{H}}} \partial_y \tilde{\phi}(\hat{v}, y, x) $ as a function of $ \hat{v} $. In particular, we focus on the case where the initial data vanish to quadratic order on the horizon, i.e. $j = 2$. Using Eq.~(\ref{eq:firstj}) from the main text, we can then estimate the late-time behavior. Indeed, one finds that
\begin{equation}
\phi(\tilde{v},\rho=0,0)=\frac{\left(\tilde{v}-1\right)^2}{216 \sqrt{\pi } \tilde{v}^{3/2}}\int_0^{+\infty}\,{\rm d}k\,\frac{e^{-\frac{k^2}{4}} k^2 \left(3-k^2\right) \tilde{v}^{-\frac{k}{2 \sqrt{3}}}}{1+\frac{k}{3 \sqrt{3}}}
\end{equation}
where we chose $\tilde{v}_0=1$, and used Eq.~(\ref{eq:gaussian2}). This integral can be evaluated exactly in terms of an incomplete Faddeeva function. However, for our purposes, it suffices to extract the first two leading terms in $\log \tilde{v}$, which can be obtained using a systematic saddle-point approximation:
\begin{equation}
\phi(\tilde{v},\rho=0,0)\approx \phi^{(2)}_{\rm NH}(\tilde{v})=\frac{2 (\tilde{v}-1)^2}{\sqrt{3 \pi }\,\tilde{v}^{3/2} \log^3\tilde{v}}\left[1-\frac{2}{\log \tilde{v}}+\mathcal{O}(\log^{-2}\tilde{v})\right]\,.
\end{equation}
At late times, the behavior of our Gaussian pulse is expected to coincide with the profile shown above. However, both the value of $\tilde{v}$ at which this agreement begins and the corresponding relative amplitude of the mode are \emph{a priori} unknown. To account for this, we again fit our numerical data to
\begin{equation}
a_0\,\phi^{(2)}_{\rm NH}\!\left(\frac{\hat{v}}{\hat{v}_0}\right)\,,
\label{eq:fit2}
\end{equation}
where $ a_0 $ and $ \hat{v}_0$ are fitting parameters.

\begin{figure}[ht]
\centering
\includegraphics[width=0.6\textwidth]{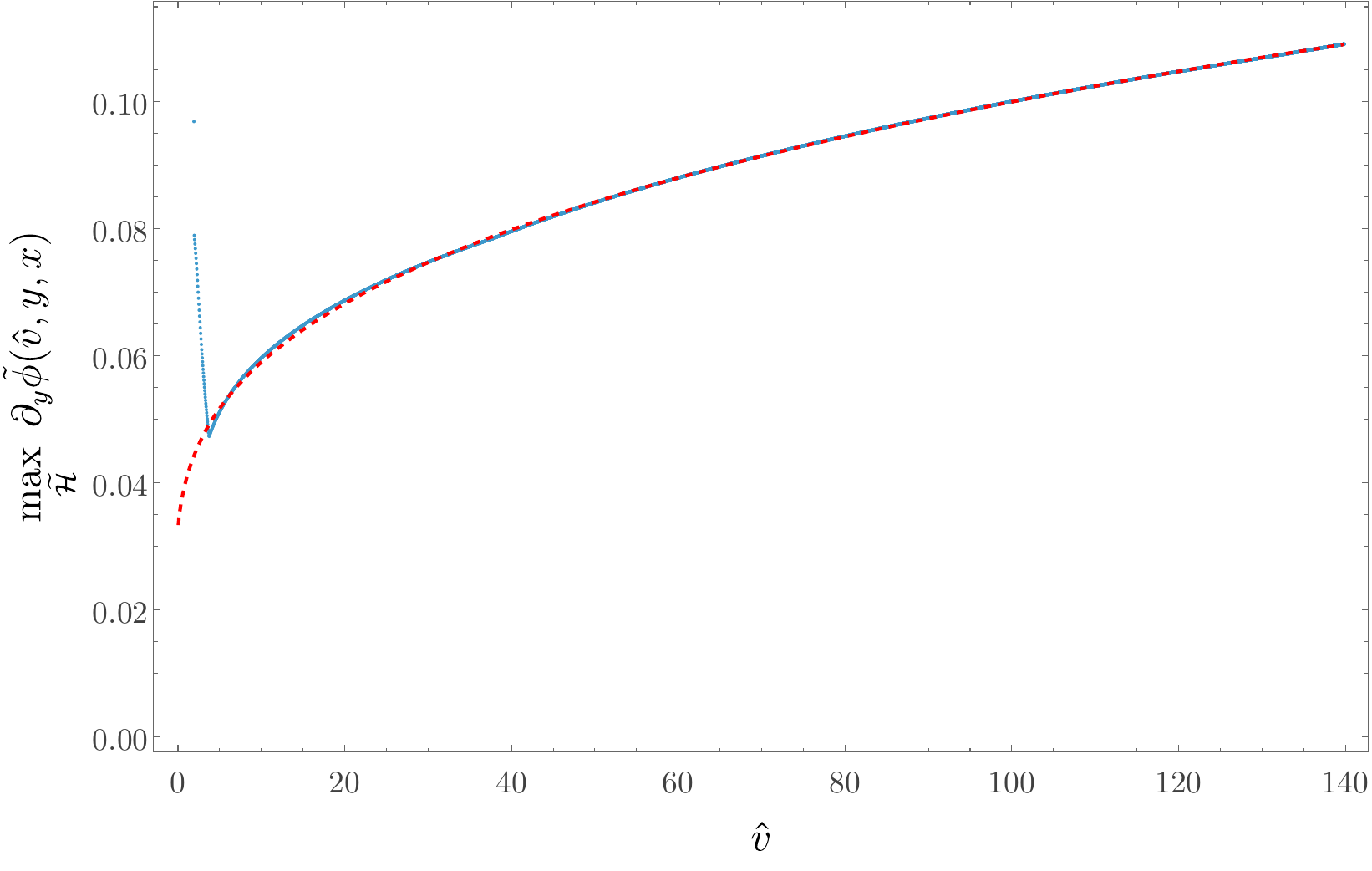}
\caption{Plot of $ \max_{\widetilde{\mathcal{H}}} \partial_y\tilde{\phi}(\hat{v},y,x)$ as a function of $v$ for the initial data specified in Eq.~(\ref{eq:initialdatascalar}) with $j=2$. The blue disks represent the numerical data, while the dashed red curve shows the $\chi^2$ best-fit to the analytic solution \eqref{eq:fit2} with parameters $a_0 \approx 934204$ and $\hat{v}_0 \approx 0.00117365$. Excellent agreement is observed for $\hat{v} \gtrsim 20$.}
\label{fig:scalar1}
\end{figure}

In Fig.~\ref{fig:scalar1}, we plot $\max_{\widetilde{\mathcal{H}}} \partial_y\tilde{\phi}(\hat{v},y,x)$ as a function of $\hat{v}$ for the initial data given in Eq.~(\ref{eq:initialdatascalar}) with $j=2$. The numerical data are shown as blue disks, while the dashed red curve represents a $\chi^2$ best-fit with $a_0 \approx 0.934204$ and $\hat{v}_0 \approx 0.00117365$. The agreement is excellent for $\hat{v} \gtrsim 20$.

\section{General linear perturbations in arbitrary dimension\label{app:B}}

In this Appendix we perform the mode analysis for linearized Einstein-Maxwell theory in arbitrary dimension $D = n+2$. Moreover, we will consider perturbations with general spatial dependence, without assuming any symmetry. 
We wish to reduce our problem to the system of decoupled equations of the form \eqref{eq:wavelike}. This was done in \cite{Jansen:2019wag}. We will quickly summarize their findings. We will denote the metric and Maxwell perturbations by $h_{\mu \nu}$ and $a_\mu$ respectively.

One may start by performing a Fourier transform with respect to the plane directions. We will denote the direction along the momentum as $x$. Then, at fixed momentum $k$, one may further decompose modes with regard to their properties under $O(n)$ rotations. There will be three sectors: tensor, vector and scalar perturbations\footnote{Note that in the main text, we could restrict ourselves to the scalar perturbations due to the assumption that the $SO(3)$ symmetry is preserved. In general, we will see the scalar sector always leads to the largest effects. As such, our late-time analysis should be valid even beyond this symmetry.}.  Within each sector, the equations of motion are written in terms of gauge-invariant combinations of perturbations. Then these are written in terms of master scalars satisfying equations of the form \eqref{eq:wavelike}\footnote{More generally, one could encounter a system of coupled wave-like equations. However, it was shown in \cite{Jansen:2019wag} that in pure Einstein-Maxwell theory, these equations can be decoupled.}.

Tensor perturbations (present only for $n>2$) are automatically gauge-invariant. They are given by $h_{yz}$ (where $y,z$ are two directions orthogonal to each other and to the momentum). The appropriate master scalar satisfies \eqref{eq:wavelike} with
\begin{equation}
    W(r) = \frac{k^2}{r^2}.
\end{equation}
Thus, they behave as a massless scalar from the point of view of the two-dimensional spacetime.
It follows that perturbations with small $k$ will vanish along the horizon slightly faster than $1/\tilde{v}$. In particular, the second radial derivative of these perturbations will actually grow with time (slightly slower than linearly). One may notice that the component $C_{\rho y \rho z}$ of the Weyl tensor will be proportional to $h_{yz,\rho \rho}$. Thus, it follows that there are null tidal forces growing along the horizon.

Vector perturbations can be written using two master scalars with different potentials. These are slightly more complicated but for our current purposes, we do not need to know what they are in the full spacetime, we only need to know their horizon values. These read:
\begin{equation}
    \mu^2_{\pm}(k) = \frac{1}{2 r_+^2 L^2}\left[
    \sqrt{2k^2 L^2 + n(n+1) {r_+^2}} \pm r_+\sqrt{ n(n+1)}
    \right]^2.
\end{equation}
We see in particular that $\mu_\pm^2 \ge 0$ (with equality only for $k=0$ mode) and so it is going to behave similarly to the tensor modes.

Scalar perturbations can be written using two master scalars with different potentials. These are much more complicated but for the current purposes, we again only need to know their horizon values. These are:
\begin{equation}
    \mu^2_{\pm}(k) = \frac{1}{r_+^2}\left[k^2+\frac{r_+^2}{L^2} n(1+n) \pm \frac{r_+}{L} \sqrt{
    4k^2 (n^2-1) + n^2 (1+n)^2 \frac{r_+^2}{L^2}}\right].
\end{equation}
$\mu_+$ is a growing function of $k$ with minimum at $k=0$:
\begin{equation}
    \mu_+^2 (k=0) = \frac{2n(1+n)}{L^2}.
\end{equation}
This corresponds to $\alpha = 2$ and so $+$ modes are quickly decaying along the horizon. $-$ modes are more interesting. In particular,
\begin{equation}
    \mu_-^2(k=0) = 0,
\end{equation}
which corresponds to $\alpha = 1$. However, for $n \ge 3$, this is not a minimum of $\mu_-$. Instead, we have
\begin{subequations}
    \begin{equation}
        \min \mu_-^2 = -\frac{(n-2)^2 (n+1)}{4(n-1)L^2}< 0
    \end{equation}
    at
    \begin{equation}
        k = \frac{r_+}{L} \sqrt{\frac{ (n-2) (n+1) (3 n-2)}{4(n-1)}},
    \end{equation}
    which corresponds to
    \begin{equation}
        \alpha = \frac{1}{2} \left(\sqrt{\frac{3 n-4}{n^2-n}}+1\right)< 1.
    \end{equation}
\end{subequations}%
It follows that the master scalar for the $-$ scalar sector decays slower. In particular, its first derivative will blow up along the horizon. However, this is not a huge difference unless the dimension is very large. As we have seen, for $n=3$, we have $\alpha \approx 0.956435$. Nevertheless, it means that at late times, we should expect the perturbations to be localized around this particular momentum. We already saw in Sec. \ref{sec:nonlinear} that this expectation is fulfilled even in the non-linear theory, validating the mode analysis. We will see in a moment that this behavior means that the tidal forces will grow as $\tilde{v}^{2-\alpha}$, in particular faster that linearly for $\alpha <1$. This implies that the tidal forces integrated along null radial geodesics will actually grow with time.\footnote{On the other hand, when $n=2$, one can check that the tidal forces at the horizon still grow but their integral along null radial geodesics actually decays. That means that even though there is a large curvature at the horizon, the observer could still cross it without too much pain. In higher dimensions, the change in the momentum of particles building up the observer will be significant. As such, crossing the horizon at late times could be deadly.} Notice that $\alpha>0$, irrespectively of the dimension. As a result, when the curvature becomes Planckian, it is contained to a sub-Planckian distance away from the horizon.

All that is left to do is to show that these results for master scalars indeed implies the aforementioned results for curvature. Let us start by defining more precisely what we mean by tidal forces. It is easy to check that 
\begin{equation}
    O_{ij} = C_{\rho i \rho j} - \frac{1}{n} \delta^{kl} C_{\rho k \rho l} \delta_{ij}
\end{equation}
is a gauge-invariant  quantity in linearized gravity since it vanishes in the background. In the gauge in which we work in the main text, $\delta^{kl} C_{\rho k \rho l} = 0$ and so $O_{ij} = C_{\rho i \rho j}$. $O_{ij}$ enters the geodesic deviation equation for null geodesics and as such provides us with a notion of tidal forces. Note that, due to the traceless nature of $O_{ij}$, it describes only the shear of a null family of geodesics\footnote{Due to the planar symmetry of the background, the radial null geodesics are clearly shear-free there.}. Since $O_{ij}$ are gauge invariant, they can be expressed in terms of the gauge-invariant quantities of \cite{Jansen:2019wag}. We find
\begin{equation}
    O_{xx} = k^2  e^{i k x} \frac{n-1}{2n} \frac{f(r)^2 \left[\mathfrak{h}_{rr}(t,r)-2 \mathfrak{h}_{rx}^{(0,1)}(t,r)\right]-f(r) \left[\mathfrak{h}_{tr}(t,r)-2 \mathfrak{h}_{rx}^{(1,0)}(t,r)\right]+\mathfrak{h}_{tt}(t,r)}{f(r)^2}.
\end{equation}
and $O_{yy} = O_{zz} =...= - \frac{1}{n-1} O_{xx}$ for all other diagonal components and all other components vanish due to the symmetry. Notice that even though $O_{ij}$ are defined in the Eddington-Finkelstein coordinates, the expression above is written in Schwarzschild-like coordinates (and thus may have problems at the horizon). We should rewrite it in terms of Eddington-Finkelstein coordinates. This should include not only a change of the coordinates but also of gauge-invariants (following \cite{Jansen:2019wag}, we will denote them by the superscript EF) to make sure that these are well-defined on the horizon. In these terms, the answer is even simpler:
\begin{equation}
    O_{xx} = \frac{(n-1)k^2  e^{i k x}}{2n} \left( \mathfrak{h}_{rr}^{\textrm{EF}}(v,r) - 2 \mathfrak{h}_{rx}^{\textrm{EF}(0,1)}(v,r) \right).
\end{equation}
Since $O_{xx}$ can be written in terms of gauge-invariant quantities, it also has a representation in terms of master scalars. The exact expression however is not very illuminating and so we will skip it here. Instead, we will write $O_{xx}$ at the horizon. It reads
\begin{equation}
    O_{xx}(r=r_+) = - \frac{k e^{ikx} (n-1)}{n} \left[
\frac{k}{r^{n-2}} \partial_r^2 \left(
\Phi_{2,0} r^{n} \right) + 2n \sqrt{n^2-1} \frac{r_+}{L} \frac{\partial_r \left(r^{n/2} \Phi_{1,0} \right)}{r^{n/2-1}}
\right]
\end{equation}
where all terms are evaluated at $r=r_+$. As promised, the tidal forces depend linearly on the second radial derivative of $\Phi_{2,0}$ and thus they grow as $v^{2-\alpha}$.

\section{\label{app:pol} The polynomials $\mathit{z}_i(r)$ used in section \ref{subsec:em}}
\begin{equation}
\mathit{z}(r)=k^2 r^4+12 M r^2-9 Q^2\,,
\end{equation}
\begin{multline}
\mathit{z}_1(r)=6 L^4 \Big[36 M Q^2 r^4 \left(k^2 r^2+4 M\right)-3 Q^4 \left(5 k^2 r^4+27 M r^2\right)
\\
M r^6 \left(k^4 r^4-24 k^2 M r^2-144 M^2\right)+27 Q^6\Big]
\\
-9 L^2 r^6\left[k^4 r^8-4 Q^2 \left(k^2 r^4+54 M r^2\right)+48 M^2 r^4+99 Q^4\right]-36 r^{12} \left(k^2 r^4+9 Q^2\right)
\end{multline}
\begin{equation}
\mathit{z}_2(r)=12 r^6 \left(k^2 r^2+6 M\right)+L^2 \left(k^4 r^6-12 k^2 M r^4+3 k^2 Q^2 r^2-144 M^2 r^2+72 M Q^2\right)
\end{equation}
\begin{multline}
\mathit{z}_3(r)=36 r^{12} \left(5 k^2 r^4+36 M r^2-9 Q^2\right)-3 L^4 \Big[3 Q^2 \left(k^4 r^8+92 k^2 M r^6+720 M^2 r^4\right)
\\
-6 Q^4 \left(19 k^2 r^4+225 M r^2\right)+2 M r^6 \left(k^4 r^4-96 k^2 M r^2-720 M^2\right)+351Q^6\Big]
\\
-9 L^2 r^6 \Big[2 r^4 \left(7 k^2 Q^2+72 M^2\right)-3 k^4 r^8+8 k^2 M r^6+216 M Q^2 r^2-171 Q^4\Big]
\end{multline}
\begin{equation}
\mathit{z}_4(r)=3 (60 M r^6-27 Q^2 r^4 + 7 k^2 r^8) + 
 L^2 \Big[3 k^2 Q^2 r^2-216 M^2 r^2 + k^4 r^6 + 18 M (7 Q^2 - k^2 r^4)\Big]
\end{equation}
\begin{equation}
\mathit{z}_5(r)=L^2 \left[Q^2 \left(7 k^2 r^4+6 M r^2\right)-6 M r^4 \left(3 k^2 r^2+4 M\right)+9 Q^4\right]+13 k^2 r^{10}+60 M r^8-45 Q^2 r^6
\end{equation}
\begin{multline}
\mathit{z}_6(r)=9 r^7 \left(7 k^2 r^4+36 M r^2-27 Q^2\right)+3 L^2 \Big[2 k^4 r^9-33 Q^2 \left(k^2 r^5+18 M r^3\right)
\\
+30 k^2 M r^7+360 M^2 r^5+243 Q^4 r\Big]
\end{multline}
\begin{multline}
\mathit{z}_7(r)=L^4 \left[6 M r^4 \left(3 k^2 r^2+28 M\right)-Q^2 \left(13 k^2 r^4+222 M r^2\right)+81 Q^4\right]
\\
-L^2 r^6 \left(k^2 r^4+60 M r^2-9 Q^2\right)+36 r^{12}
\end{multline}
\begin{equation}
\mathit{z}_8(r)=L^2 \left(k^2 r^6+6 M r^4\right)-9 r^8
\end{equation}
\begin{multline}
\mathit{z}_9(r)=9 \left(3 k^2 r^{10}+20 M r^8-15 Q^2 r^6\right)+L^2 \Big[9 Q^2 \left(3 k^2 r^4+2 M r^2\right)
\\
-2 \left(k^4 r^8+27 k^2 M r^6+36 M^2 r^4\right)+27 Q^4\Big]
\end{multline}

\section{\label{app:lagrange}Spectral collocation methods in unbounded domains - using Laguerre functions}
We define the Laguerre functions
\begin{equation}
\tilde{L}^{\alpha}_n(x)=e^{-\frac{x}{2}}L_n^{\alpha}(x)\,,
\end{equation}
in terms of the generalised Laguerre polynomials $L_n^{\alpha}(x)$ of order $n$.

We then determine the extrema of the generalised Laguerre polynomials of order $n=N+1$ via
\begin{equation}
\partial_xL_{N+1}^{\alpha}(x)=0\,,
\end{equation}
which can be shown to have exactly $N$ zeroes, which we label as $y^\alpha_k$, with $k=1,\ldots,N$. We then construct the set $\{x^\alpha_k\}$ defined as $x^\alpha_0=0$, $x^\alpha_{k}=y^\alpha_k$, with $k=1,\ldots,N$.

The weights $w^\alpha_k$ are built as follows
\begin{equation}
w^\alpha_k=\left\{
\begin{array}{c}
\frac{(\alpha+1)\Gamma(\alpha+1)^2\Gamma(N+1)}{\Gamma(N+\alpha+2)}\quad\text{if}\quad k=0
\\
\\
\frac{\Gamma(N+\alpha+1)}{N!(N+\alpha+1)}\frac{1}{\tilde{L}^\alpha_N(x^\alpha_{k})^2}\quad\text{for}\quad k=1,\ldots,N
\end{array}
\right.
\end{equation}

The synthesis matrix (taking from modal space to physical space) is given by
\begin{equation}
S^{\alpha}_{ik}=\tilde{L}^{\alpha}_i(x_k),\quad\text{with}\quad i,k =0,\ldots,N\,.
\end{equation}
Note that if we regard $S^{\alpha}_{ik}$ as the components of a matrix $\mathbf{S}^{\alpha}$, $\mathbf{S}^{\alpha}$ is an $(N+1)\times(N+1)$ matrix.

Additionally, the analysis matrix (taking from physical space to modal space) is defined by
\begin{equation}
A^\alpha_{ji}=\frac{j!}{\Gamma(j+\alpha+1)}w^{\alpha}_i S^\alpha_{ij},\quad\text{with}\quad i,j=0,\ldots,N
\end{equation}
Again, if we regard $A^{\alpha}_{ik}$ as the components of a matrix $\mathbf{A}^{\alpha}$, $\mathbf{A}^{\alpha}$ is an $(N+1)\times(N+1)$ matrix.

To differentiate, we simply introduce the matrix $\mathbb{D}$, whose components we define as
\begin{equation}
\mathbb{D}_{ki}=\left\{
\begin{array}{c}
0,\quad\text{if}\quad i<k
\\
-\frac{1}{2},\quad\text{if}\quad i=k
\\
-1,\quad\text{if}\quad i>k
\end{array}\right.\quad\text{with}\quad k,i=0,\ldots,N\,.
\end{equation}
and build the differentiation matrix (acting solely on physical space)
\begin{equation}
\mathbf{D}_1^{\alpha}=\mathbf{S}^\alpha\mathbb{D} \mathbf{A}^\alpha\,,
\end{equation}
where in the above standard matrix multiplication is assumed. For higher derivatives along $x$, we simply take
\begin{equation}
\mathbf{D}_p^{\alpha}=\mathbf{S}^\alpha\underbrace{\mathbb{D}\ldots\mathbb{D}}_{p} \mathbf{A}^\alpha\,,
\end{equation}

The above construction yields well-conditioned matrices up to about $N = 1000$, which suffices for our purposes. See \cite{shen2000stable} for further details. The solutions under consideration exhibit power-law decay in $1/x$, potentially modulated by highly oscillatory integrals. For problems of this type, the convergence of standard numerical methods is typically algebraic in $1/N$, sometimes accompanied by logarithmic corrections. Fortunately, in our case, the primary numerical limitation arises from the development of large \emph{radial} gradients. These gradients necessitate a local increase in the number of elements near the horizon, which ultimately slows the simulation due to the Courant–Friedrichs–Lewy condition.

\section{\label{app:appbessel}Hankel transform}

Take a function $f(R)$, with $R>0$, and define the following integral transform
\begin{equation}
\hat{f}(k)=\frac{2}{\pi}\int_0^{+\infty}f(R)\mathbb{Y}_k(R)R^2{\rm d}R\,.
\end{equation}
Then, using the properties of Bessel functions, it follows that
\begin{equation}
f(R)=\int_0^{+\infty}\hat{f}(k)\mathbb{Y}_k(R)k^2{\rm d}k\,.
\end{equation}

We now turn to the numerical implementation of this procedure. 
Fortunately, the required integral has been derived in \cite{ChouinardBaddour2017}; 
our task is simply to adapt it with a few modifications to suit the present problem.

We begin by introducing a cutoff radius $R_{\max}$. Let $N \in \mathbb{N}$ be the number of grid points, such that the integration domain is discretized into $N-1$ points. The corresponding $R$- and $k$-grids are defined as
\begin{equation}
R_i = \frac{i R_{\max}}{N},
\qquad
k_i = \frac{i \pi}{R_{\max}},
\qquad
i = 1, \ldots, N-1 \, .
\end{equation}

The discrete transform can then be written in terms of the matrix
\begin{equation}
\mathbf{M}_{ij} = 
\frac{R_{\max}^2}{N^2}\,
\frac{2^{3/2}}{\pi^{5/2}}\,
\frac{1}{\sqrt{k_i}}\,
\frac{J_{1/2}\!\left(\tfrac{i j \pi}{N}\right)}{J_{3/2}(j \pi)}\,
\sqrt{R_j}\, .
\end{equation}

Given a function $f(R)$, we sample it on the radial grid as 
$f_i \equiv f(R_i)$. The corresponding transform coefficients are then obtained as
\begin{equation}
\hat{f}_j \equiv \hat{f}(k_j) \;\approx\; 
\left( \mathbf{M} \cdot \mathbf{f} \right)_j \, .
\end{equation}

\bibliography{planarads}{}
\bibliographystyle{utphys-modified}

\end{document}